\documentclass[12pt]{article}
\usepackage{amsmath, amsthm, amssymb}
\usepackage{graphicx,psfrag,epsf}
\usepackage{subfig}
\usepackage{enumerate}
\usepackage{natbib}
\usepackage{bm}
\usepackage[T1]{fontenc}

\usepackage{hyperref}
\hypersetup{
    colorlinks,
    linktocpage,
    citecolor=blue,
    filecolor=blue,
    linkcolor=blue,
    urlcolor=blue
}

\DeclareMathOperator*{\argmax}{arg\,max}

\newcommand{\pkg}[1]{{\normalfont\fontseries{b}\selectfont #1}}
\let\proglang=\textsf
\let\code=\texttt

\newcommand{\yb}{\bm{y}}

\newcommand{\Gig}[1]{\text{GIG}\!\left(#1\right)}

\newcommand{\Gammad}[1]{ \mathcal{G}\!\left(#1\right)}
\newcommand{\Normal}[1]{\mathcal{N}\!\left(#1\right)}

\newcommand{\Normult}[2]{\mathcal{N}_{#1}\!\left(#2\right)}

\newcommand{\Betad}[1]{\mathcal{B}\!\left(#1\right)}

\newcommand{\ym}{{\bm{ y}}}             
\newcommand{\allindex}{\bullet}
\newcommand{\ymi}[1]{\ym_{#1,\allindex}}
\newcommand{\fm}{{\bm{ f}}}             
\newcommand{\fmj}[1]{{\bm{f}}_{#1,\allindex}}             
\newcommand{\Vary}{{\bm{\vary}}}         
\newcommand{\vary}{\Sigma}         
\newcommand{\dimy}{m}                     
\newcommand{\load}{\Lambda}                 
\newcommand{\parstar}[1]{#1 ^\star} 
\newcommand{\facloadstar}{\parstar{\facload}} 
\newcommand{\facload}{\bm{\load}} 
\newcommand{\facrow}[1]{\facload_{#1,\allindex}}
\newcommand{\faccol}[1]{\facload_{\allindex, #1}}
\newcommand{\facstarcol}[1]{\facloadstar_{\allindex, #1}}
\newcommand{\fac}{f}                      
\newcommand{\facm}{{\bm{ \fac}}}        
\newcommand{\facmi}[1]{\facm_{#1,\allindex}} 
\newcommand{\facmstar}{\parstar{\facm}} 
\newcommand{\facmstari}[1]{\parstar{\facm}_{#1,\allindex}} 
\newcommand{\errorz}{\epsilon}             
\newcommand{\error}{\varepsilon}             
\newcommand{\errorstate}{\zeta}             
\newcommand{\errorm}{\bm{\error}} 
\newcommand{\errorzm}{\bm{\errorz}} 
\newcommand{\errorstatem}{\bm{\errorstate}} 
\newcommand{\Vare}{{\bm{\vare}}}      
\newcommand{\vare}{{{U}}}      
\newcommand{\hsv}{h}             
\newcommand{\hsvm}{{\bm{\hsv}}}             
\newcommand{\hsvmi}[1]{{\bm{\hsv}}_{#1,\allindex}}             
\newcommand{\hsvmstari}[1]{\parstar{\hsvm}_{#1,\allindex}} 
 
\newcommand{\Dm}{{\bm{ D}}}             
\newcommand{\Vm}{{\bm{ V}}}             

\newcommand{\mupartrue}{\mupar_\text{true}}
\newcommand{\phipartrue}{\phipar_\text{true}}
\newcommand{\sigmapartrue}{\sigma_\text{true}}

\newcommand{\mupar}{\mu}
\newcommand{\phipar}{\phi}
\newcommand{\sigmapar}{\sigma}
\newcommand{\etat}{\eta}
\newcommand{\etatm}{\bm{\etat}}

\newcommand{\Diag}[1]{\mbox{\text diag}\!\left(#1\right)} 
\newcommand{\dimmat}[2]{#1\times #2}  
\newcommand{\bfz}{{\bm{{0}}}}         
\newcommand{\identm}{{\bm{I}}}       
\newcommand{\identy}[1]{{\identm}_{#1}} 
\newcommand{\unit}[1]{{\identm}_{#1}} 
\newcommand{\trans}[1]{#1 ^{'}}         

\newcommand{\facloadtrue}{\bm{\Lambda}_\text{true}} 
\newcommand{\nfactrue}{r}                     
\newcommand{\new}{^\text{new}}
\newcommand{\old}{^\text{old}}
\renewcommand{\new}{^{\mbox{\rm \tiny new}}}
\renewcommand{\old}{^{\mbox{\rm \tiny old}}}

\theoremstyle{plain}

\theoremstyle{definition}
\newtheorem{alg}{Algorithm}

\newcommand{\e}{e}

\graphicspath{{figures/}}

\begin{document}

\def\spacingset#1{\renewcommand{\baselinestretch}%
{#1}\small\normalsize} \spacingset{1}

\newpage 

  \title{\bf Efficient Bayesian Inference for Multivariate Factor Stochastic Volatility Models}
 
  \author{
   \normalsize \textsc{Gregor Kastner}\\
   \normalsize \textit{Department of Finance, Accounting and Statistics}\\
  \normalsize  \textit{WU Vienna University of Economics and Business, Austria}\\
 \normalsize  \url{gregor.kastner@wu.ac.at} \\[.5em]
 \normalsize \textsc{Sylvia Fr\"uhwirth-Schnatter}\\
 \normalsize \textit{Department of Finance, Accounting and Statistics}\\
 \normalsize \textit{WU Vienna University of Economics and Business, Austria}\\
 \normalsize \url{sfruehwi@wu.ac.at} \\[.5em]
 \normalsize \textsc{Hedibert Freitas Lopes}\\
 \normalsize \textit{Insper, Brazil}\\
 \normalsize \url{hedibertfl@insper.edu.br}
}


\maketitle

\begin{center}
 \bf
Forthcoming in:\\
Journal of Computational and Graphical Statistics\\
\href{https://dx.doi.org/10.1080/10618600.2017.1322091}{doi:10.1080/10618600.2017.1322091}
\end{center}

\smallskip
\begin{abstract}
We discuss efficient Bayesian estimation of dynamic covariance matrices in multivariate time series through a factor stochastic volatility model. In particular, we propose two interweaving strategies \citep{yu-men:cen} to substantially accelerate convergence and mixing of standard MCMC approaches. Similar to marginal data augmentation techniques, the proposed acceleration procedures exploit non-identifiability issues which frequently arise in factor models. Our new interweaving strategies are easy to implement and come at almost no extra computational cost; nevertheless, they can boost estimation efficiency by several orders of magnitude as is shown in extensive simulation studies. To conclude, the application of our algorithm to a 26-dimensional exchange rate data set illustrates the superior performance of the new approach for real-world data.
\end{abstract}

\noindent%
\vfill

\thispagestyle{empty}

\newpage
\spacingset{1.45} 

\section{Introduction}

The analysis of multivariate time series has become a vivid research area over the last decades, where both methodological as well as computational advances have made it possible to estimate more and more complex models. In parallel, real-world applications with an ever-increasing amount of data call for the joint modeling of many simultaneous and often co-varying observations over time. However, already the number of pair-wise co-movements increases quadratically with the number of time series, let alone higher-dimensional dependency structures. This property, often referred to as the \emph{curse of dimensionality},
can often be mitigated in various ways by imposing a lower-dimensional latent factor structure, thereby effectively reducing the number of parameters to a feasible amount. In the paper at hand, we particularly focus on the case where these factors are allowed to have time-varying variances which in turn drive the multivariate dynamics. To the best of our knowledge, models of this type have first been discussed by \citet{jac-etal:bayJBES}, \citet{she:sta}, and \citet{kim-etal:sto}. We particularly focus on the model formulation brought forward by \citet{chi-etal:ana}.

Applications of multivariate factor stochastic volatility models typically reside in the field of financial econometrics, most prominently in areas that involve accurate quantification of uncertainty and risk. Examples thereof are asset allocation \citep[e.g.][]{agu-wes:bay, han:ass, zho-etal:bay} and asset pricing \citep[e.g.][]{nar-scr:bay}. These models extend standard factor pricing models such as the arbitrage pricing theory \citep{ros:arb} and the capital asset pricing model \citep{sha:cap,lin:val} by relaxing the assumption that the multivariate volatility dynamics are constant over time.

Statistical estimation of these models can be challenging, and a variety of solutions such as quasi-maximum likelihood \citep[e.g.][]{har-etal:mul} or simulated maximum likelihood \citep[e.g.][]{lie-ric:cla,jun-koo:msv} have been proposed. For medium to high dimensional problems, Bayesian MCMC estimation \citep{pit-she:tim, agu-wes:bay, chi-etal:ana, han:ass, omo-etal:sto} is probably the most efficient estimation method, however, it is associated with a considerable computational burden when the number of assets is moderate to large.

The aim of this work is to outline a reliable method for Bayesian inference that performs well for a wide range of data sets while at the same time being easy to implement and convenient to extend.
Therefore, we combine an efficient method for estimating univariate stochastic volatility models
introduced by \citet{kas-fru:anc}
 with a standard Gibbs sampler for regression problems. To ensure fast convergence and proper mixing of the MCMC chains we augment this simple procedure with interweaving strategies introduced by \citet{yu-men:cen}. Through extensive simulation studies and a real-world example, we demonstrate the effectiveness of our procedure which can boost sampling efficiency by a factor of 100 and more.

The remainder of this paper is structured as follows. Section~\ref{sec:mod} establishes notation for the factor stochastic volatility model framework and discusses questions about model specification and identification. Section~\ref{sec:est} gives an in-depth exposure to the estimation algorithm and its implementation, whereby the focus is placed on the novel interweaving strategies employed. Section~\ref{sec:sim} presents measures of sampling efficiency for simulated data sets and compares the algorithms presented. Section~\ref{sec:app} discusses a case study with $26$ daily EUR exchange rates. Section~\ref{sec:con} concludes.

\section{The Multivariate Factor Stochastic Volatility Model} \label{sec:mod}

In a multivariate framework, the quadratic growth of the number of covariances alongside their inherent time-variability calls for a model which is sufficiently parsimoniously specified. At the same time, the model needs to be flexible enough to have the potential to capture typical features of financial and economic time series such as volatility clustering and volatility co-movement. On top of that, common irregularities in the data require the model to be robust with respect to idiosyncratic shocks.

The multivariate factor stochastic volatility (SV) model \citep{chi-etal:ana} aims at uniting simplicity with flexibility and robustness. It is \emph{simple} in the sense that the potentially high-dimensional observation space is reduced to a lower-dimensional orthogonal latent factor space, just like in the case of the classic factor model. It is \emph{flexible} in the sense that these factors are allowed to exhibit volatility clustering, and it is \emph{robust} in the sense that idiosyncratic deviations are themselves stochastic volatility processes, thereby allowing for the degree of volatility co-movement to be time-varying.

\subsection{Model Specification}

For each point in time $t=1,\ldots,T$, let $\ym_t=\trans{(y_{1t}, \ldots, y_{\dimy t})}$ be a zero-mean vector of $m$ observed returns and let
$\fm_t=\trans{(f_{1t}, \ldots,  f_{\nfactrue t})}$ be a vector of $\nfactrue$ unobserved latent factors. In analogy to the static factor model, the observations are assumed to be driven by the latent factors and the idiosyncratic innovations. In the case of the factor  stochastic volatility model, however, both  the idiosyncratic innovations as well as the latent factors  are allowed to have time-varying variances, depending on $m+\nfactrue$ latent  volatilities
$\hsvm_t=(\hsvm_t^U,\hsvm_t^V)$, where $\hsvm_t^U= \trans{(\hsv_{1t}, \ldots,  \hsv_{m t})}$ and $\hsvm_t^V= \trans{(\hsv_{m+1,t}, \ldots,  \hsv_{m+\nfactrue, t})}$.
 In short, we have
\begin{eqnarray}  \label{fac1}
\ym_t =  \facload \facm_t + \Vare_t (\hsvm_t^U)^{1/2} \errorzm_t, \qquad  \facm_t = \Vm_t (\hsvm^V _t) ^{1/2} \errorstatem_t,
\end{eqnarray}
%
where $\facload$ is an unknown $\dimmat{\dimy}{\nfactrue}$ factor loadings matrix,
$\Vare_t (\hsvm^U_t)=\Diag{\exp(\hsv_{1t}),\ldots,\exp(\hsv_{\dimy t})}$ is a diagonal $\dimy \times \dimy$ matrix containing the idiosyncratic (series-specific) variances, and
 $\Vm_t (\hsvm^V_t)=\Diag{\exp(\hsv_{\dimy+1,t}),\ldots,\exp(\hsv_{\dimy+\nfactrue, t})}$ is a diagonal $\nfactrue \times \nfactrue$ matrix containing the factor variances. These variances are themselves modeled as latent variables whose logarithms follow independent autoregressive processes of order one, i.e.\ for $i=1,\ldots,m+r$:
\begin{eqnarray}  \label{fac3}
\hsv_{it} &=& \mupar_i +  \phipar_i (\hsv_{i,t-1}- \mupar_i )  + \sigmapar_i \etat_{it},
\end{eqnarray}
with unknown initial value $\hsv_{i0}$.

All innovations are assumed to follow independent standard normal distributions, i.e.\ $\errorzm_t \sim \Normult{\dimy}{\bfz,\unit{\dimy}}$, $\errorstatem_t \sim    \Normult{\nfactrue}{\bfz,\unit{\nfactrue}}$, and $\etatm_{t} \sim \Normult{\dimy + \nfactrue}{\bfz,\unit{\dimy+\nfactrue}}$, where $\etatm_{t} = (\etat_{1t}, \dots, \etat_{m+r,t})'$. This implies following structure:
\begin{eqnarray}  \label{fac4}
\ym_t = \facload \facm_t +  \errorm_t,  \qquad \facm_t|\hsvm_t \sim  \Normult{\nfactrue}{\bfz,\Vm_t (\hsvm^V_t)},
\end{eqnarray}
with $\errorm_t|\hsvm_t \sim \Normult{\dimy}{\bfz,\Vare_t (\hsvm_t)}$. 
One of the main reasons for estimating a factor SV model is to reliably estimate the potentially time-varying conditional covariance matrix   of $\ym_t$  which, for the model at hand, is  given by $\text{cov}(\ym_t|\hsvm_t) = \Vary_t(\hsvm_t) = \facload \Vm_t(\hsvm^V_t) \facload' + \Vare_t(\hsvm^U_t)$.
Note that because $\Vare_t (\hsvm^U_t)$ is diagonal, all covariances between the component series are governed by the latent factors. Marginally with respect to $\hsvm_t$, $\ym_t$ is a process with non-Gaussian stationary distribution.




\subsection{Identification Issues}
\label{sec:ident}

Whenever certain combinations of parameter values result in (almost) identical maxima in the likelihood function, estimation of the corresponding parameter values from data can become impossible. Consequently, observationally equivalent parameter constellations must be ruled out for reliable statistical inference and a large body of literature dealing with this issue has arisen. In particular, \citet{fru-lop:par} give an overview of recent advances in the context of static Bayesian factor models, and \citet{sen-fio:ide} specifically discuss identification for models where the factors exhibit conditional heteroscedasticity (but innovations are assumed to be homoscedastic). If not dealt with properly, usually through certain restrictions on the parameter space, sensible interpretation of the posterior distribution is not possible (``nonidentifiability''). In less severe cases (``near-nonidentifiability''), MCMC algorithms and other estimation procedures often lack convergence and thus provide unreliable results. For the model at hand, we face several issues related to this problem.

First, to prevent factor rotation and column switching, one option is to follow the usual convention and set the upper triangular part of $\facload$ to zero and $\Diag{\facload}$ nonzero \citep[e.g.][]{gew-zho:mea}. Doing so, however, imposes an -- often unwanted -- order dependence. We therefore also discuss the possibility to leave the factor loadings matrix unrestricted and deal with column switching through post-processing of the MCMC draws.

Second, without identifying the scaling of either the $j$th column  of  $\facload$ or the variance of  $\fac_{jt}$, the model is not identified. The usual remedy \citep[e.g.][]{agu-wes:bay, chi-etal:ana, han:ass, lop-car:fac, nak-wes:dyn, zho-etal:bay} is that the diagonal loading elements in model~(\ref{fac4}) are fixed to one, i.e.\ $\load_{jj}=1$, for $j=1, \ldots, \nfactrue$, while the level $\mupar_{m+j}$  of the factor volatilities $\hsv_{m+j,t}$ in model (\ref{fac3}) (which corresponds to the scaling  of $\fac_{jt}$) is modeled to be unknown. This approach implies that the first $\nfactrue$ variables are leading the factors and thus makes variable ordering an even more important modeling decision. To alleviate this issue, we leave the diagonal elements $\load_{jj}$ in model~(\ref{fac4}) unrestricted, an intuitive interpretation being that ``leadership'' of a factor can be  shared by several series.
Instead, we fix the level  $\mupar_{m+j}$ of the factor volatilities $\hsv_{m+j,t}$ at zero:
\begin{eqnarray}
&\hsv_{it}= (1-\phipar_i)\mupar_i +  \phipar_i \hsv_{i,t-1} + \sigmapar_i \etat_{it}, \quad i=1, \ldots,m,&
\nonumber \\
& \hsv_{m+j,t}= \phipar_{m+j} \hsv_{m+j,t-1} + \sigmapar_{m+j} \etat_{m+j,t}, \quad j=1, \ldots, r.& \label{fac3fix}
\end{eqnarray}
This assumption, alongside the prior distribution on the loadings introduced in Section~\ref{sec:pri}, identifies the factor variance. 

Finally, each column of $\facload$ is only identified up to a possible sign switch. We deal with this (lightweight) identification issue a posteriori, meaning that we run our MCMC sampler in the unrestricted model and identify signs afterwards, see Section~\ref{alg:sign} in Appendix~\ref{app:mcmc} for details.

Factor model (\ref{fac4}) together with the $m+r$ SV models (\ref{fac3fix}) defines our baseline parameterization, however alternative parameterizations will be exploited in Section~\ref{alg:asis} in the context of efficient MCMC estimation of the factor SV  model.

\section{Bayesian Inference} \label{sec:est}

We perform Bayesian inference based on a set of carefully selected proper priors which are introduced in Section~\ref{sec:pri} and develop efficient schemes for full conditional MCMC sampling in the remaining subsections.

\subsection{Prior Distributions} \label{sec:pri}

Independently for each $i \in \{1, \dots, m+r\}$, priors for the univariate SV processes are chosen  as in \citet{kas-fru:anc}:
$p(\mu_i,\phi_i,\sigma_i)$ = $p(\mu_i)p(\phi_i)p(\sigma_i)$, where the level $\mupar_i \in \mathbb{R}$ is equipped with the usual normal prior $\mu_i \sim \Normal{b_{\mu}, B_{\mu}}$, the persistence parameter $\phipar_i \in (-1,1)$ is chosen according to $(\phi_i+1)/2 \sim \Betad{a_0, b_0}$ as in \citet{kim-etal:sto}, and the volatility of log variance $\sigmapar_i \in \mathbb{R}^+$ is implied by $\sigma_i^2 \sim B_\sigma\times \chi^2_1=\Gammad{\frac{1}{2},\frac{1}{2B_\sigma}}$. The initial state $\hsv_{i0}$ is distributed according to the stationary distribution of the AR($1$) process (\ref{fac3}), i.e.\ $\hsv_{i0}|\mupar_i,\phipar_i,\sigmapar_i \sim \Normal{\mupar_i,\sigmapar_i^2/(1-\phipar_i^2)}$.
For  every unrestricted  element of the factor loadings matrix we choose independent zero-mean Gaussian distributions, i.e.\ $\load_{ij} \sim \Normal{0, B_\load}$.

\subsection{Full Conditional MCMC estimation} \label{sec:ful}

Bayesian inference operates directly in the latent variable model (\ref{fac4}) and (\ref{fac3fix}) and relies on data augmentation by introducing the latent volatilities $\hsvm=\{\hsvmi{i}\}, i=1,\ldots, \dimy+\nfactrue$, where $\hsvmi{i}=(\hsv_{i0}, \hsv_{i1},\dots, \hsv_{iT})'$,
 and the latent factors $\fm=\{\fmj{j}\}, j=1,\ldots, \nfactrue$,
where $\fmj{j}=(\fac_{j1},\dots, \fac_{jT})'$, as latent data. 
This allows to set up a simple scheme for full conditional MCMC sampling  which is outlined in Algorithm~\ref{facsvalg} and discussed in detail thereafter.

\begin{alg} \label{facsvalg}
 Choose
 appropriate starting values for
 $\mu_i$, $i \in \{1,\dots,m\}$,
 $\phi_i$ and $\sigma_i$, $i \in \{1,\dots,m+r\}$,
 as well as $\facload$, $\hsvm$ and $\fm$ and repeat the following steps:
 \begin{enumerate}
  \item[(a)] Perform in total  $m+r$ univariate SV updates of the $m$  idiosyncratic variances  $\hsvmi{i}$
as well as  the parameters  $(\mupar_i, \phipar_i,\sigmapar_i)$,  independently  for each $i=1,\dots,m$,
    and of  the $r$  factor variances $\hsvmi{m+j}$
  as well as the parameters $(\phipar_{m+j},\sigmapar_{m+j})$,  independently  for each  $j=1,\dots,r$.

 \item[(b)] For $i=1,\dots,m$, sample each row $\facrow{i}$ of the factor loading matrix from $\facrow{i}|\facm,\ymi{i},\hsvmi{i}$.  This step constitutes $m$ independent  $\tilde r_i$-variate regression problems with $T$ observations, where $\tilde r_i$ denotes the number of unrestricted elements in $\facrow{i}$.

\item[(b*)] Redraw the diagonal elements of $\facload$ through interweaving into the state equation for the latent factors (\emph{shallow interweaving}) or through interweaving into the state equation for the latent volatilities (\emph{deep interweaving}).

\item[(c)] For $t=1,\dots,T$, sample $\facm_t$ from $\facm_t|\facload,\ym_{ t},\hsvm_{ t}$, constituting $T$ independent $r$-variate regression problems with $m$ observations. 
\end{enumerate}
\end{alg}

\noindent For Step~(a), observe that conditional on knowing the latent factors $\fm$ and the loadings $\facload$, we are dealing with $ \dimy+\nfactrue$ independent, univariate SV models where the latent state equations (\ref{fac3fix}) are combined with following observation equations:
\begin{eqnarray}
 \log (y_{it} -  \facrow{i} \facm_t)^2  &=&  \hsv_{it} + \log  \errorz_{it}^2 , \qquad i=1,\ldots,\dimy, \label{facob1} \\
 \log \fac_{jt} ^2  &=&  \hsv_{m+j,t} + \log  \errorstate_{jt}^2, \qquad j=1,\ldots, \nfactrue. \label{facob2}
 \end{eqnarray}
Hence, sampling the latent volatilities
 $\hsvmi{i}$ as well as the parameters $(\mupar_i, \phipar_i,\sigmapar_i)$ for  $ i=1,\ldots,\dimy+\nfactrue$ (with $\mupar_i=0$ for $i>\dimy$) in Step~(a) amounts to $m+r$ univariate SV updates.  Consequently, the substantial amount of research on this matter which has emerged in the last two decades can directly be applied. In particular,
 we follow recent findings in \citet{kas-fru:anc}, where an efficient sampling scheme is proposed and evaluated, and simply use the implementation in the \proglang{R} package \pkg{stochvol} \citep{kas:dea} as a ``plug-in'' for Step~(a) of the factor SV sampler presented in Algorithm~\ref{facsvalg}; see Appendix~\ref{alg:univ} for more details and additional references on MCMC estimation for univariate SV models.

On the other hand, conditional on knowing the latent volatilities $\hsvm$, we are dealing in (\ref{fac4}) with a factor model with heteroscedastic errors. Nevertheless, given $\hsvm$,  $\fm$ and $\facload $ may be sampled conditionally on each other from the respective multivariate normal distributions in a similar manner  as for a standard factor model \citep{lop-wes:bay}. This approach is conceptually straightforward, see  Appendix~\ref{alg:facload} for details how to sample in Step~(b) each row $\facrow{i}$ of the factor loading matrix  from $\facrow{i}|\facm,\ymi{i},\hsvmi{i}$, where $\ymi{i}=(y_{i1},\ldots, y_{iT})'$, and Appendix~\ref{alg:samfac} for details how to sample  in Step~(c) the factor  $\facm_t$   from $\facm_t|\facload,\ym_{ t},\hsvm_{ t}$ for $t=1,\dots,T$.

After discarding a certain amount of initial draws (the \emph{burn-in}), the {\it standard full conditional sampler} iterates steps~(a), (b) and (c) of Algorithm~\ref{facsvalg}, but not (b*), and should, in principle,  yield draws from the joint posterior distribution.
However, when estimating factor SV models through such an MCMC scheme, slow convergence and poor mixing (i.e.\ high correlation of posterior draws) can become a potentially prohibitive issue.
This phenomenon substantiates in enormous autocorrelation of posterior draws -- even after thinning -- and can render MCMC output practically useless.
 For certain data sets,  the burn-in phase may take extremely long and a huge amount of samples has to be discarded before the draws can be considered to emerge from the posterior distribution. Additionally, even after burn-in, these draws often show extraordinarily high autocorrelation and thus only explore the target distribution painstakingly slowly. These so-called \emph{badly mixing} samplers do not only prolong computation time, they also frequently lead to unreliable estimates and misleading results. The simulation study in Section~\ref{sec:sim} illustrates that this can happen for the standard full conditional sampler even with data simulated from the true model, see e.g.\ the top of the two panels in Figure~\ref{simstudy:trace}. Consequently, a carefully crafted posterior simulator is of utmost importance.

To overcome this problem, \citet{chi-etal:ana} propose to sample the factor loading matrix $\facload$ from the marginalized conditional posterior $p(\facload|\ym,\hsvm)$, without conditioning on the factors $\facm$. This distribution, however, is not available in closed form, and to sample from it requires a rather involved Metropolis-Hastings update where the proposal distribution is based on numerically maximizing the often high-dimensional conditional likelihood function and approximating its Hessian matrix at every MCMC iteration. To avoid this potential bottleneck, we employ the simpler full conditional procedure outlined in Algorithm~\ref{facsvalg} but enhance it in Step~(b*) by employing two variants of an ancillarity-sufficiency interweaving strategy (ASIS) \citep{yu-men:cen}, called
{\it shallow interweaving} and {\it deep interweaving}, which are explained in detail in
Section~\ref{alg:asis}.

Applications to simulated data in Section~\ref{sec:sim}  as well as to exchange rate data in Section~\ref{sec:app} illustrate how adding Step~(b*) boosts  MCMC dramatically, in particular for deep interweaving; compare e.g.\ the top panel in Figure~\ref{simstudy:trace} to the remaining panels.
Section~\ref{sec:impl} in Appendix~\ref{app:mcmc} provides comments on practical implementation of the boosted Algorithm~\ref{facsvalg} using the 
\proglang{R} package \pkg{factorstochvol} \citep{kas:fac}.

\subsection{Boosting Full Conditional MCMC through Interweaving} \label{alg:asis}

As discussed in Section~\ref{sec:ful}, the {\it standard full conditional sampler} outlined in Algorithm~\ref{facsvalg} is based on data augmentation  in the parameterization (\ref{fac4}) and (\ref{fac3fix})  of the  factor  SV model and suffers from slow convergence like so many other  MCMC schemes which alternate between sampling from the full conditionals of the latent states and the model parameters. A large literature has emerged discussing various techniques to improve such algorithms, in particular
reparameterization \citep{pap-etal:gen}, marginal data augmentation \citep{van-men:art}, and interweaving strategies \citep{yu-men:cen}.

Reparameterization relies on data augmentation in a different parameterization of the model with alternative latent variables. In particular, so-called non-centered parameterizations where unknown model parameters are moved from the latent state equation to the observation equation proved to be useful, see e.g.~\citet{fru-wag:sto} in the context of state space modeling of time series.
However, MCMC estimation based on different data augmentation schemes will often be efficient in separate regions of the parameter space, as demonstrated  e.g.~by \citet{kas-fru:anc} in the context of univariate SV models. This suggests to combine different data augmentation schemes to obtain an improved sampler.

Marginal data augmentation employs a randomly sampled ``working parameter'' to transform the baseline parameterization to an expanded, unidentified latent variable model in which the model parameters are updated conditional on the (randomly) transformed latent variables. This technique has been applied to the basic factor model, using the undefined scaling of the factors as a working parameter  \citep{gho-dun:def,fru-lop:par},  however, it is not easily extended to factor SV models, in particular if the latent volatilities should be part of the acceleration scheme.

 The \emph{ancillarity-sufficiency interweaving strategy (ASIS)}, introduced by \citet{yu-men:cen}, provides another principled way to interweave two different data augmentation  schemes by  re-sampling certain parameters conditional on the latent variables in an alternative parameterization of the model, thereby combining ``best of different worlds''. ASIS has been successfully employed in a variety of contexts such as  univariate SV models \citep{kas-fru:anc} and
dynamic linear state space models \citep{sim:app,sim-etal:int}.
To  boost  Algorithm~\ref{facsvalg},  we apply ASIS  to the factor SV model in the present paper.  Two interweaving strategies -- called \emph{shallow interweaving} and \emph{deep interweaving} -- are derived in Section~\ref{boostsub}, where  the diagonal elements $\load_{11}, \ldots,  \load_{\nfactrue \nfactrue}$ of the factor loadings matrix are resampled in Step~(b*) in two alternative parameterizations of the model.

As will become clear in the following sections,
 deep interweaving typically yields the highest sampling efficiency gains and is thus the generally recommended strategy. However, also shallow interweaving
 has its merits. First, being conditionally conjugate, it is somewhat easier to implement. Second, it can be applied also to static factor models which are by
  construction not suited for deep interweaving \citep[see also][]{bit-fru:ach}.

\subsubsection{Shallow and Deep Interweaving}     \label{boostsub}

As discussed in Section~\ref{sec:ident}, our baseline parameterization  (\ref{fac4}) and (\ref{fac3fix})   is just one of several alternative ways to handle the scaling problem inherent in factor SV models and this identification issue is exploited  by our schemes.

The parameterization underlying shallow interweaving constrains the diagonal elements  of the factor loadings matrix to be equal to 1, whereas the variances of the factors depend on $\nfactrue$  unknown scaling parameters $\Dm=\Diag{\load_{11}, \ldots,  \load_{\nfactrue \nfactrue}}$.
The latent volatility processes are modeled as in the baseline parameterization (\ref{fac3fix}), whereas the factor model takes a different  form:
\begin{eqnarray}  \label{fac5}
\ym_t = \facloadstar \facmstar_t +  \errorm_t,  \qquad \facmstar_t| \hsvm_t, \load_{11}, \ldots,  \load_{\nfactrue \nfactrue}
 \sim  \Normult{\nfactrue}{\bfz,\Dm ^2 \Vm_t (\hsvm^V_t)},
\end{eqnarray}
with a lower triangular loading matrix $\facloadstar$  where $\parstar{\load}_{11}=1, \ldots,  \parstar{\load}_{\nfactrue \nfactrue}=1$. The idiosyncratic errors  $\errorm_t$ are distributed as in (\ref{fac4}).  Factor model  (\ref{fac4})  in the baseline parameterization
can be transformed into factor model (\ref{fac5}) through a simple linear transformation:
\begin{equation}
 \facmstar_t = \Dm   \facm_t, \qquad t=1,\ldots,T, \qquad
\facloadstar=\facload  \Dm ^{-1}.
\label{shallowtrans}
\end{equation}
Boosting through shallow interweaving consists of three parts. First, transformation (\ref{shallowtrans}) is used to move  the current posterior draws of the latent factors $\facm_t$ and  the factor loading matrix  $\facload$  from the baseline parameterization  to parameterization (\ref{fac5}). Second, the scale parameters $\load_{11}, \ldots,  \load_{\nfactrue \nfactrue}$, contained in $\Dm$, are resampled  in parameterization (\ref{fac5}), conditionally on the transformed values $\facmstar$, from $p(\load_{11}, \ldots,  \load_{\nfactrue \nfactrue}|\facmstar, \facloadstar, \hsvm)$.
Finally,   the new values  $\load_{11} \new, \ldots,  \load_{\nfactrue \nfactrue}\new $ are used in transformation (\ref{shallowtrans}) to move
$\facmstar_t$  and  $\facloadstar$  back to new draws  $\facm_t \new$ and  $\facload \new$  in the   baseline parameterization.

It is evident from transformation (\ref{shallowtrans}) that shallow interweaving only affects the factors and the factor loading matrix, whereas the latent volatilities remain untouched. This is the feature that makes shallow interweaving also applicable to static factor models.
However, to achieve boosting also for the $r$  factor volatilities, deep interweaving is based on an alternative SV model for the factor volatilities where the  level
 is assumed to be unknown. The parameterization underlying deep interweaving relies on the factor model
\begin{eqnarray}  \label{fac6}
\ym_t = \facloadstar \facmstar_t +  \errorm_t,  \qquad \facmstar_t| \hsvmstari{m+j}
 \sim  \Normult{\nfactrue}{\bfz, \Diag{e^{\parstar{\hsv}_{\dimy+1,t}},\ldots,e^{\parstar{\hsv}_{\dimy+\nfactrue, t}}}},
\end{eqnarray}
where $\facloadstar$ has the same structure as in the factor model (\ref{fac5}) for shallow interweaving and the idiosyncratic errors $\errorm_t$ are distributed as before, with the univariate SV models for the $m$ underlying volatilities following (\ref{fac3fix}).
However, the $r$ latent factor volatilities $\parstar{\hsv}_{m+j,t}$ follow alternative univariate SV models where the level is $\mupar_{m+j} =\log \load_{jj}^2$ rather than zero:
\begin{eqnarray}
 \parstar{\hsv}_{m+j,t}
 &=& \mupar_{m+j}(1-\phipar_{m+j}) + \phipar_{m+j}\parstar{\hsv}_{m+j,t-1} + \sigmapar_{m+j}\eta_{m+j,t}.
 \label{fac3deep}
\end{eqnarray}
This parameterization can be motivated by moving the parameters $\load_{11}, \ldots,  \load_{\nfactrue \nfactrue}$ from  factor model (\ref{fac5})  into  SV model (\ref{fac3deep}), since:
\begin{equation*}
 \parstar{f}_{jt} |\load_{jj}, \hsv_{m+j,t}
 \sim \Normal{0, \load_{jj}^2\e^{\hsv_{m+j,t}}}
 = \Normal{0, \e^{\log\load_{jj}^2+\hsv_{m+j,t}}} = \Normal{0, \e^{\parstar{\hsv}_{m+j,t}}} .
 \label{deeplik}
\end{equation*}
Hence, the baseline parameterization can
be transformed into parameterization (\ref{fac6})  and  (\ref{fac3deep}) by applying transformation (\ref{shallowtrans}) to the factors  and the factor loadings, as well as the following transformation to the factor volatilities:
  \begin{eqnarray}
\parstar{\hsv}_{m+j,t}  = \hsv_{m+j,t} + \log \load_{jj}^2, \quad t=0,\ldots , T, \quad
j=1,\ldots, r.
\label{deeptrans}
\end{eqnarray}
Boosting through deep interweaving also consists of three parts:
first, transformations (\ref{shallowtrans}) and (\ref{deeptrans}) are used to move from the current draws of 
 $\facm_t$,  
 $\facload$ and the factor log-variances $\hsv_{m+j,t}$
from the baseline parameterization to parameterization (\ref{fac6}) and (\ref{fac3deep}).
Second, the scale parameters $\load_{11}, \ldots,  \load_{\nfactrue \nfactrue}$ are
resampled in parameterization  (\ref{fac3deep})  conditionally on the transformed values $\hsvmstari{m+j}=(\parstar{\hsv}_{m+j,0}, \ldots, \parstar{\hsv}_{m+j,T})'$ from $p(\load_{11}, \ldots,  \load_{\nfactrue \nfactrue}|\hsvmstari{m+1}, \cdots, \hsvmstari{m+r}, \facloadstar)$.
Based on the new values  $\load_{11}\new, \ldots,  \load_{\nfactrue \nfactrue}\new$, transformations (\ref{shallowtrans})
  and (\ref{deeptrans}) are inverted to move  $\facmstar_t$,  $\facloadstar$, $\parstar{\hsv}_{m+j,t}$
    back to new draws  $\facm_t \new$,  $\facload \new$, $\hsv_{m+j,t} \new$ in the baseline parameterization.

Both interweaving strategies are summarized in Algorithm~\ref{StepBstar}.
Details on resampling $\load_{jj}\new$ are provided in Section~\ref{boostdetail}.
It is evident that deep interweaving affects the factors, the factor loading matrix as well as  the latent factor volatilities and for this reason is more effective in boosting MCMC for factor SV models than shallow interweaving.

\begin{alg}[\textbf{Shallow and Deep Interweaving}] \label{StepBstar}
Denote the original posterior draws for $\faccol{j}$,  $\facmi{j}$, and $\hsvmi{m+j}$ in Algorithm~\ref{facsvalg} by $\faccol{j} \old$, $\facmi{j} \old$,  and $\hsvmi{m+j}\old$ and perform following steps independently for each $j=1,\ldots, r$ in Step~(b*):
\begin{enumerate}
\item[(b*-1)]  Determine the vector $\facstarcol{j}$,
containing  the $k_j$ free parameters $\parstar{\load}_{ij}=\load_{ij}\old/\load_{jj}\old$ in the $j$th column of the transformed factor loading matrix $\facloadstar$. 

\item[(b*-2)] For {shallow  interweaving}, define $\facmstari{j}=\load_{jj}\old \facmi{j} \old$ and sample  a new value $\load_{jj}\new$ from $p(\load_{jj}|\facstarcol{j}, \facmstari{j},\hsvmi{m+j})$.
For {deep interweaving}, define $\hsvmstari{m+j}= \hsvmi{m+j}\old
    + 2 \log |\load_{jj}\old| $ and
sample   $\load_{jj}\new$ from $p(\load_{jj}|\facstarcol{j}, \hsvmstari{m+j},\phipar_{m+j}, \sigmapar_{m+j})$; see Section~\ref{boostdetail} for details.

\item[(b*-3)] Update  $\faccol{j}$,  $\facmi{j}$, and, for deep interweaving, also $\hsvmi{m+j}$:
     \begin{eqnarray*}
 \faccol{j}= \frac{\load_{jj}\new}{\load_{jj}\old } \faccol{j} \old, \qquad   \facmi{j}=\frac{\load_{jj}\old }{\load_{jj}\new} \facmi{j} \old,   \qquad
 \hsvmi{m+j}= \hsvmi{m+j}  \old  + 2 \log \left| \frac{\load_{jj}\old }{\load_{jj}\new}\right|.
\end{eqnarray*}
\end{enumerate}
\end{alg}

\subsubsection{Sampling the scaling parameters in the alternative representations}     \label{boostdetail}

To derive the full conditional posterior distribution of $\load_{jj}$, we combine the appropriate full conditional likelihood function  with the Gaussian prior  $\load_{jj} \sim \Normal{0, B_\load}$. In addition,
the prior  $\facstarcol{j}|\load_{jj}^2 \sim \Normult{k_j}{0, B_\load /\load_{jj}^{2} \unit{k_j}}$ of the transformed factor loadings in column $j$ contributes to the  posterior distribution of $\load_{jj}^2$ because its scale depends on $\load_{jj}^2$.

For shallow interweaving, we sample $\load_{jj}^2$  and define $\load_{jj}\new$
as the square root of $\load_{jj}^2$.
Combining the likelihood obtained from factor model (\ref{fac5}) with the implied prior $\load^2_{jj} \sim \Gammad{1/2, 1/(2B_\load)}$ and $p(\facstarcol{j}|\load_{jj}^2)$ yields
\begin{eqnarray*}
 p(\load_{jj}^2|\facstarcol{j}, \facmstari{j},\hsvmi{m+j})
 &\propto&
 p(\facmstari{j}|\hsvmi{m+j}, \load_{jj}^2)
  p(\facstarcol{j}|\load_{jj}^2)p(\load_{jj}^2),
\end{eqnarray*}
which is the product of $T$ univariate Gaussian densities with $\load_{jj}^2$ appearing as part of the variance, $k_j$ univariate Gaussian densities with $\load_{jj}^2$ appearing as part of the precision, and one Gamma density with $\load_{jj}^2$ appearing as argument. Thus, the resulting posterior distribution of $\load_{jj}^2$ is Generalized Inverse Gaussian, i.e.
\begin{equation}     \label{shallowpost}
  \load_{jj}^2|\facstarcol{j}, \facmstari{j},\hsvmi{m+j} \sim \Gig{\frac{1+k_j-T}{2},
  \frac{1}{B_\load}
   \left(1+(\facstarcol{j})'\facstarcol{j}
   \right),
   \sum_{t=1}^T \frac{{f_{jt}^*}^2}{\e^{\hsv_{m+j,t}}}},
  \end{equation}
where $\Gig{p,a,b}$ has a density proportional to $ x^{p-1}\exp\left\{-\frac{1}{2}(ax+b/x)\right\}$.
  Given an efficient method to draw from the GIG such as the adaptive rejection sampling algorithm provided by \citet{hoe-ley:gen}, sampling from (\ref{shallowpost}) is straightforward. For practical implementation, we use the \proglang{R} package \pkg{GIGrvg}
  \citep{r:gig} which provides a \proglang{C/C++} interface to avoid the cost of interpreting code at every MCMC iteration, thereby rendering the re-updating negligible in terms of overall computation time.

For deep interweaving, we sample $\load_{jj}$ indirectly through $\mupar_{m+j} =\log \load_{jj}^2$. Combining   the implied prior $p(\mupar_{m+j}) \propto \exp\left\{\mupar_{m+j}/2 - \e^{\mupar_{m+j}}/{(2B_\load)}\right\}$ 
with the likelihood obtained from SV model (\ref{fac3deep})
and the priors $\parstar{h}_{m+j,0}| \mupar_{m+j}, \phipar_{m+j}, \sigmapar^2_{m+j} \sim \Normal{\mupar_{m+j},\sigmapar_{m+j}^2/(1-\phipar_{m+j}^2)}$ and
$\facstarcol{j}|\mupar_{m+j} \sim \Normult{k_j}{0, B_\load \e^{-\mupar_{m+j}} \unit{k_j}}$
 yields the posterior
\begin{eqnarray*} \label{deeppost}
p(\mupar_{m+j}|\facstarcol{j}, \hsvmstari{m+j},\phipar_{m+j}, \sigmapar^2_{m+j}) 
 \propto
  p(\hsvmstari{m+j}|\mupar_{m+j}, \phipar_{m+j}, \sigmapar^2_{m+j})
  p(\facstarcol{j}| \mupar_{m+j}) p(\mupar_{m+j}),
\end{eqnarray*}
which has a non-standard form. 
To generate draws from this  density, 
we consider an independence  Metropolis-Hastings update in the spirit of \cite{kas-fru:anc}.
 Since the likelihood $p(\parstar{\hsv}_{m+j,1}, \ldots, \parstar{\hsv}_{m+j,T}  |\parstar{\hsv}_{m+j,0}, \mupar_{m+j}, \phipar_{m+j}, \sigmapar^2_{m+j})$ is the kernel of a Gaussian density in $\mupar_{m+j}$, it can be used to construct an auxiliary posterior under a conjugate auxiliary prior $p_\text{aux}(\mupar_{m+j}|\sigmapar^2_{m+j},\phipar_{m+j}) \sim \Normal{0,B_0\sigmapar^2_{m+j}/(1-\phipar_{m+j})^2}$
 with $B_0$ large.
Consequently, we draw a proposal $\mupar_{m+j}^\text{prop}$ from the $\Normal{m_j^\mu,S_j^\mu}$ distribution with:
\begin{eqnarray*}
m_j^\mu= \frac{\sum_{t=1}^{T-1} \parstar{\hsv}_{{m+j},t} +  (\parstar{\hsv}_{m+j,T}-\phipar_{m+j} \parstar{\hsv}_{m+j,0})/(1-\phipar_{m+j})}{T + 1/B_0}, \quad S_j^\mu=
\frac{\sigma_{m+j}^2/(1-\phipar_{m+j})^2}{T + 1/B_0}. 
\end{eqnarray*}
Denoting the old value of $\mupar_{m+j}$ by $\mupar_{m+j}\old$, this proposal gets accepted with probability $\min(1,R)$,
where
\[R= \frac{ p(\facstarcol{j}|\mupar^\text{prop}_{m+j})
 p(\parstar{\hsv}_{m+j,0}|\mupar_{m+j}^\text{prop}, \phipar_{m+j}, \sigmapar^2_{m+j})
 p(\mupar_{m+j}^\text{prop})}
 { p(\facstarcol{j}| \mupar\old_{m+j})
 p(\parstar{\hsv}_{m+j,0}|\mupar_{m+j}\old, \phipar_{m+j}, \sigmapar^2_{m+j})
 p(\mupar_{m+j}\old)
}
\times
\frac{p_\text{aux}(\mupar_{m+j}\old|\sigmapar^2_{m+j},\phipar_{m+j})}
{p_\text{aux}(\mupar_{m+j}^\text{prop}|\sigmapar^2_{m+j},\phipar_{m+j})}.\]
In case of acceptance, set $\load_{jj}^\text{new} = \e^{\mupar_{m+j}^\text{prop}/2}$; otherwise, let $\load_{jj}^\text{new} = \load_{jj}^\text{old}$. 

To conclude, we add two remarks. First, note that it is easy to combine both interweaving schemes within the MCMC sampler by daisy-chaining the corresponding steps. Second, note that any nonzero element of the $j$th factor column $\faccol{j}$ can be used to boost its mixing (not only the diagonal element $\load_{jj}$). This is useful in particular when no loading matrix restrictions are enforced as it cannot be guaranteed that the diagonal elements are nonzero. Thus, in such situations, one could use a randomly selected (nonzero) element of each loadings column instead. Alternatively, one could also use the element whose absolute value is maximal.

\section{Simulation Study} \label{sec:sim}

In order to compare the different algorithms in terms of sampling efficiency, a simple simulation experiment is conducted. We use $m=10$ (simulated) series and $r=2$ (simulated) factors to generate $T=1000$ observations, thereby imposing the usual lower triangular constraint. The data generating parameter values -- listed in Table~\ref{simstudy:truevals} in Appendix~\ref{app:sim} -- are kept constant, whereas the data generating process as well as the estimation procedure is repeated $100$ times. Each time, the draws are initialized at the data generating values; then, $5100\,000$ draws are obtained of which $100\,000$ are discarded as burn-in.
Prior hyperparameters are set as follows: $B_\load = 1$, $b_\mu = 0$, $B_\mu = 100$, $a_0 = 20$, $b_0 = 1.5$, and $B_\sigma = 1$.

\begin{figure}[t]
 \centering
 \includegraphics[width = .49\textwidth, page=1]{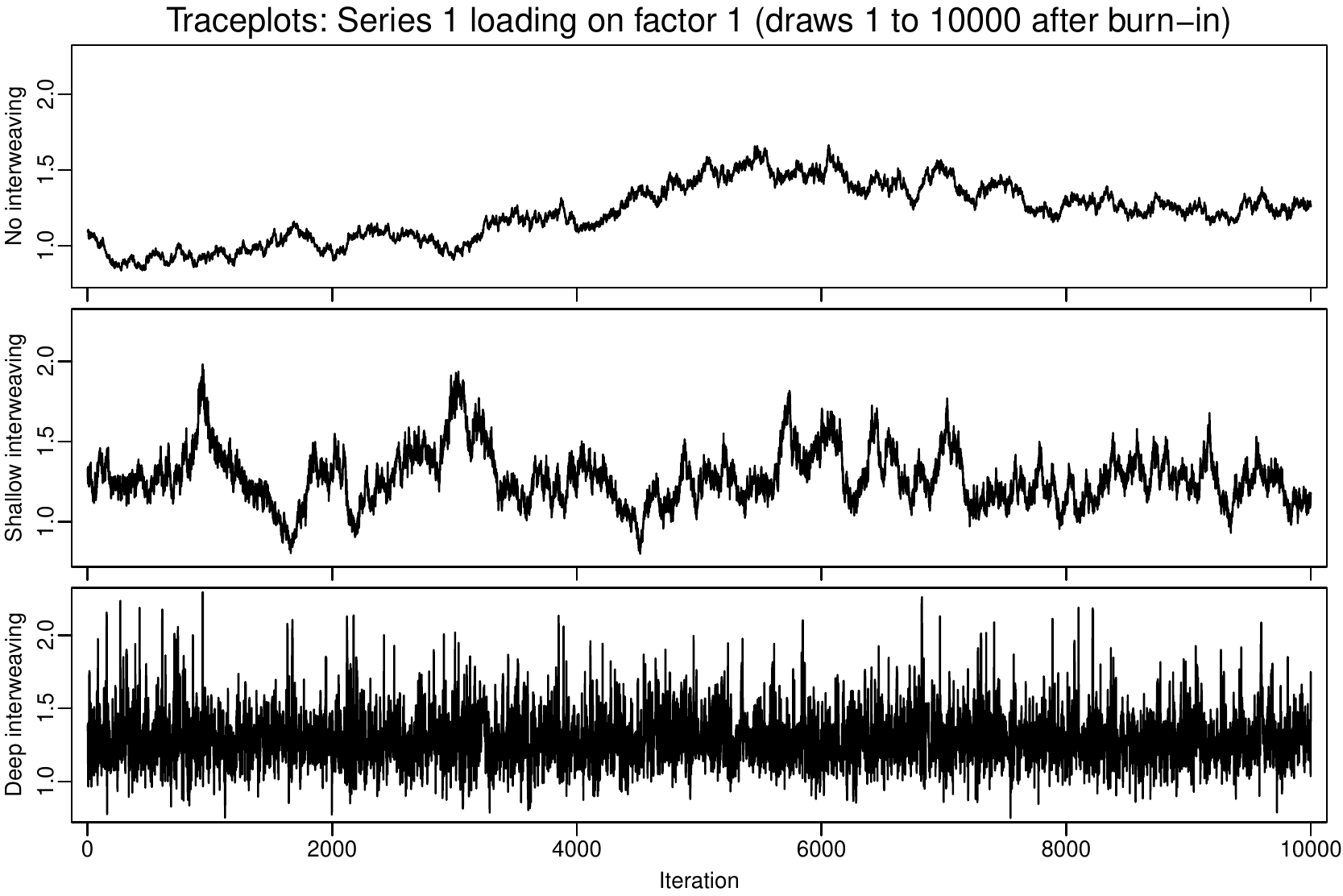}
 \includegraphics[width = .49\textwidth, page=1]{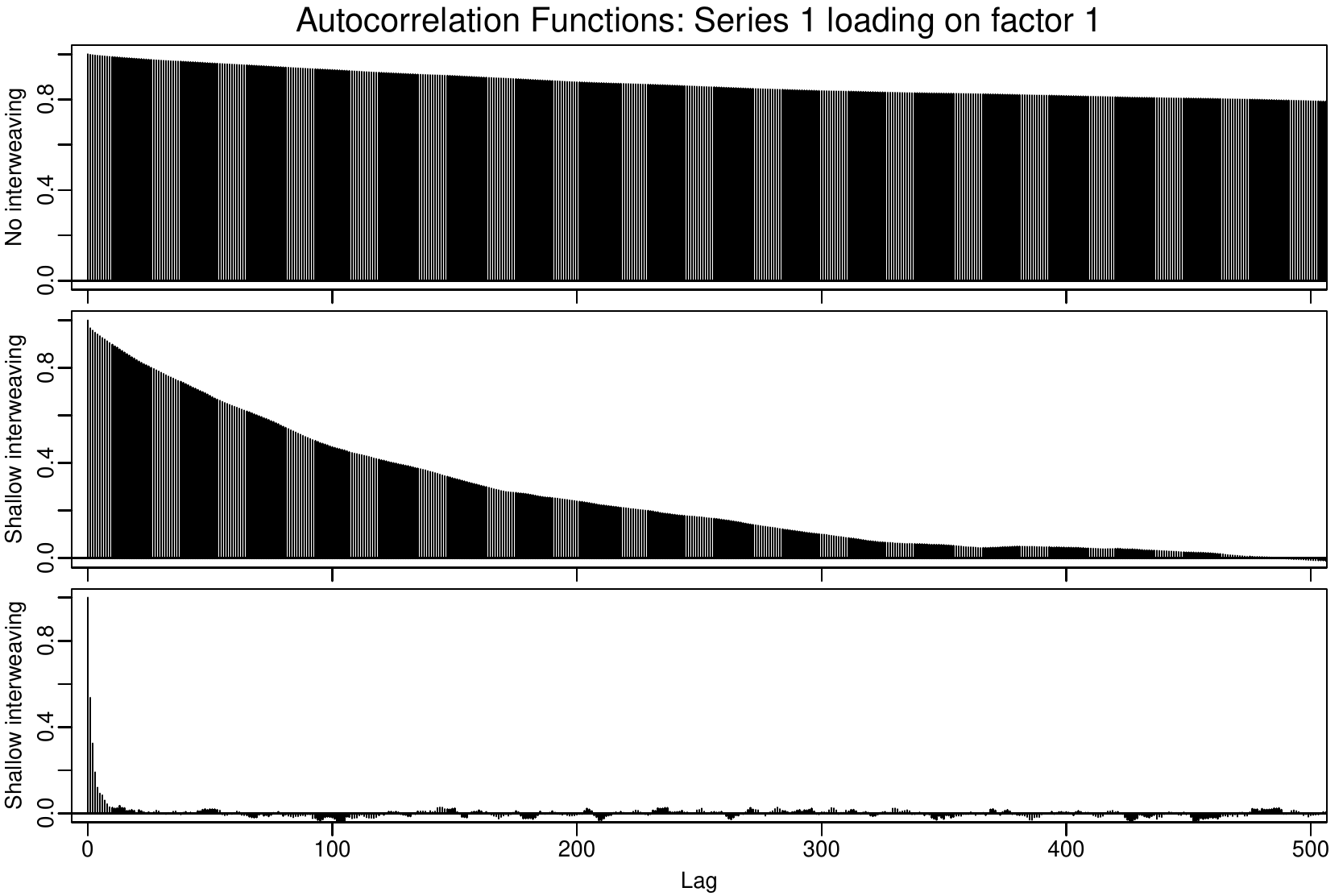}
\caption{Trace plots of  $10\,000$ draws from $p(\load_{11}|\yb)$ (left hand side) and empirical autocorrelation functions of all $5000\,000$ draws (right hand side) obtained via the standard sampler (top), shallow interweaving (middle), and deep interweaving (bottom).} \label{simstudy:trace}
\end{figure}

To gain insight about the mixing behavior of the different sampling strategies, trace plots (i.e.\ time series plots of the MCMC draws) for $\load_{11}$ are displayed in the left hand panel of Figure~\ref{simstudy:trace}. 
  Even though the plots depict only the first $10\,000$ iterations after burn-in, it becomes very clear that the mixing of the non-interwoven sampler is extremely slow. The algorithm doesn't seem to explore the posterior distribution within a reasonable amount of draws which renders this output practically useless in terms of posterior inference. Moreover, the burn-in period for this sampler would need to be chosen extremely long to avoid strong dependence on the starting values. This situation is slightly mitigated when using shallow interweaving; nevertheless, mixing is still poor and for reliable posterior inference many draws are required. Turning towards the deeply interwoven sampler, one can observe quick mixing and hardly any visible autocorrelation.


Investigating autocorrelations of the draws via the empirical autocorrelation function confirms this picture; the right hand panel of Figure~\ref{simstudy:trace} shows that the empirical autocorrelation function for draws from $p(\load_{11}|\yb)$ decays very quickly for the sampler using deep interweaving which is not the case for the other two samplers, where visible autocorrelation remains even at large lags.


A convenient and common way of measuring sampling (in)efficiency is by means of the \emph{inefficiency factor (IF)}, sometimes called the \emph{(integrated) autocorrelation time}. It is defined as the ratio of the numerical variance of a statistic which is estimated from the Markov chain to the variance of that statistic when estimated from independent draws, thereby quantifying the relative loss of efficiency when inferring from correlated as opposed to independent samples. In other words, to achieve the same inferential accuracy about some posterior moment of some parameter as with $k$ independent samples, $\text{IF}\times k$ MCMC draws are required. For the paper at hand, we use the \proglang{R} package \pkg{coda}~\citep{plu-etal:cod} to estimate the inefficiency factors.

\begin{figure}[t]
 \centering
\includegraphics[width = .4\textwidth]{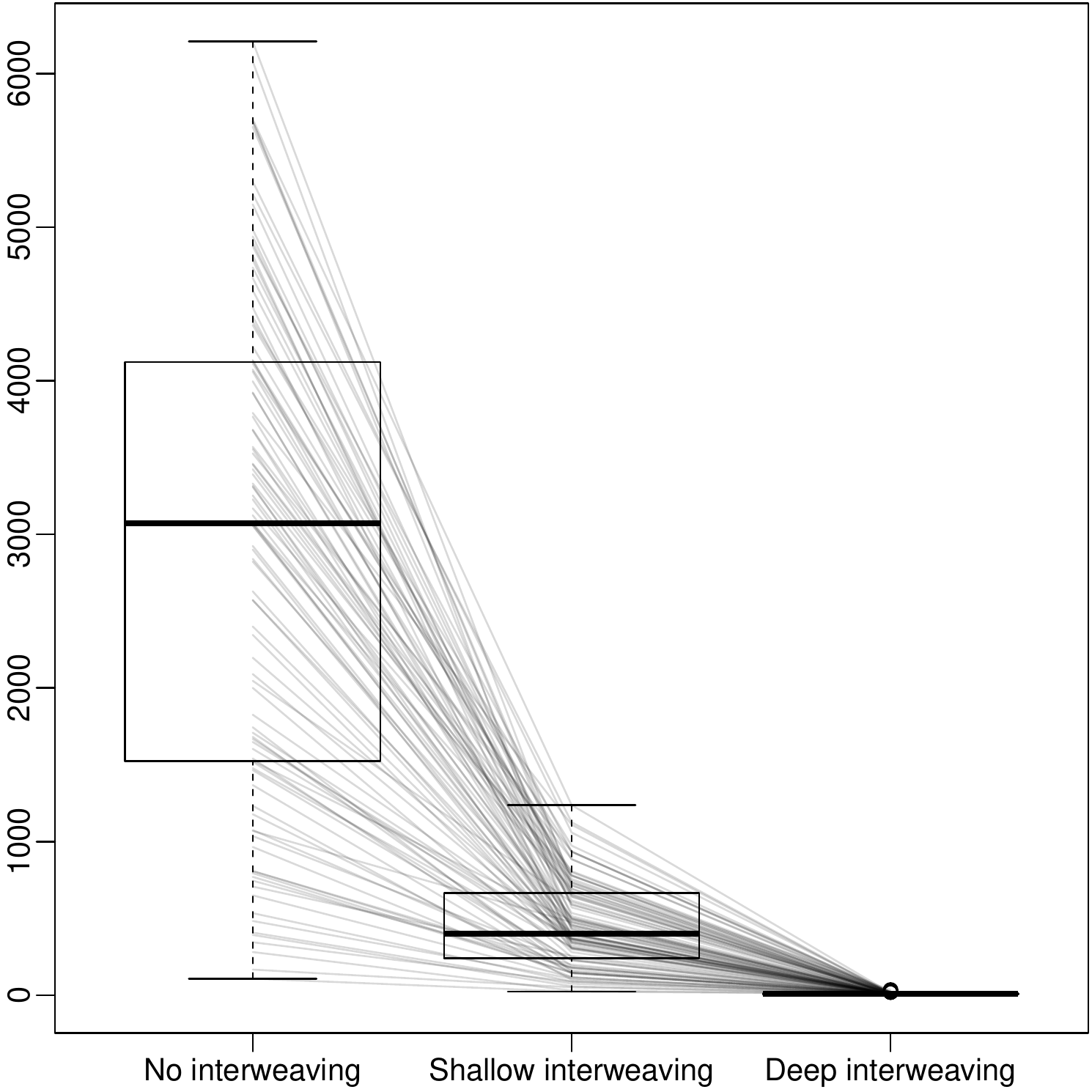}
\includegraphics[width = .42\textwidth]{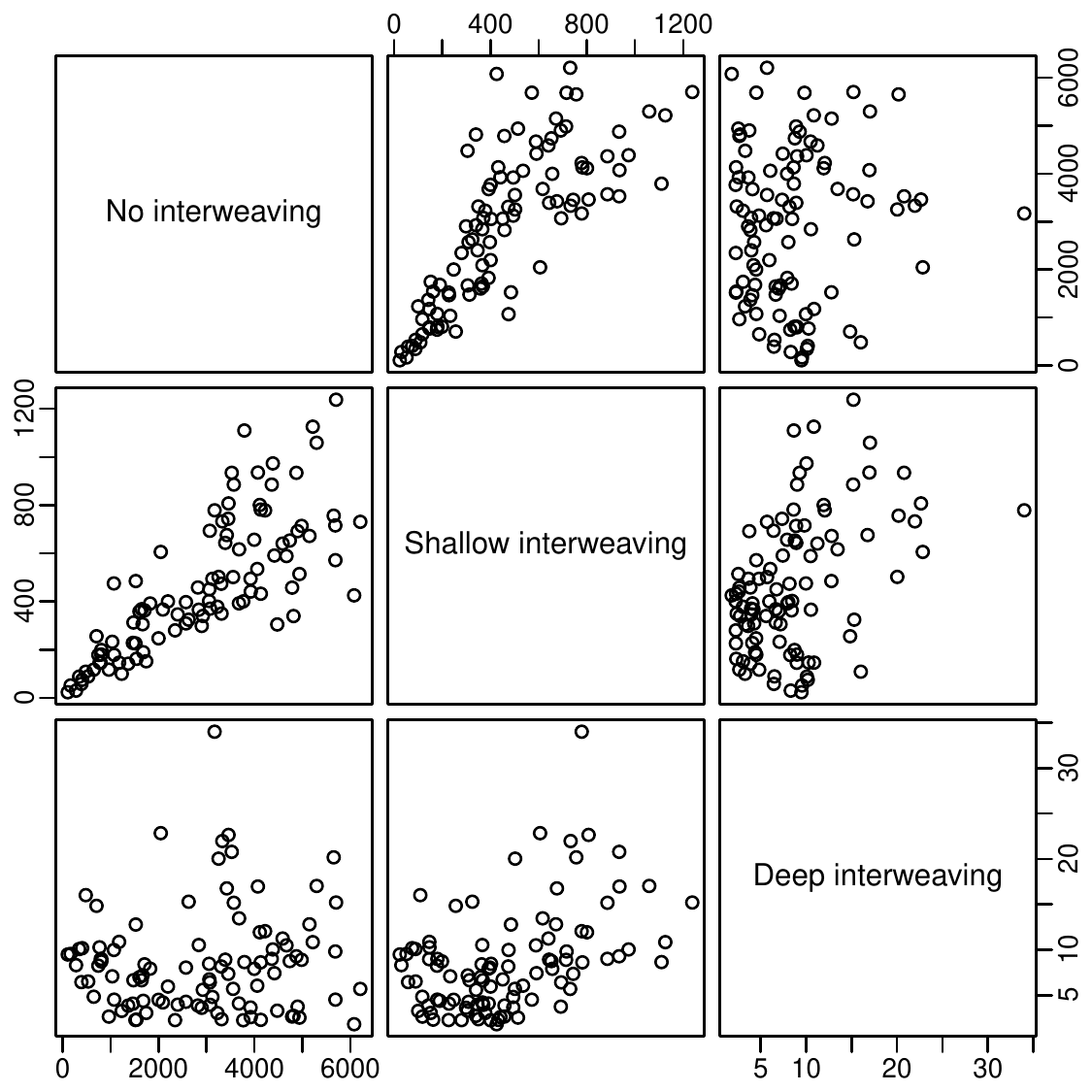}
\caption{Left: Boxplots of estimated inefficiency factors for posterior draws from $p(\load_{11}|\yb^{[i]})$ where $\yb^{[i]}, i \in \{1,\dots,100\}$, denote artificially generated data sets whose underlying parameters are identical, cf.\ Table~\ref{simstudy:truevals} in Appendix~\ref{app:sim}. Right: Pairwise scatter plots thereof.}
 \label{simstudy:ifbox}
\end{figure}

Moreover, when investigating performance of MCMC samplers through simulation studies, it is of great importance to take sample variation into account; even when identical parameter values are used for the generation of latent variables and data, sampling (in)efficiency may vary greatly. To illustrate this, we show box plots of the inefficiency factors stemming from repeated data generating processes in the left panel of Figure~\ref{simstudy:ifbox}. Note the enormous range for the standard sampler; depending on the data, IFs of $5000$ or more are not uncommon, while at the same time IFs of around $100$ can be observed. However, independently of the actual data, interweaving attenuates this effect drastically and increases efficiency uniformly. The right panel of Figure~\ref{simstudy:ifbox} shows pairwise scatter plots of these IFs. Note that shallow interweaving yields efficiency improvements which are more or less independent of the actual data (around five-fold for all data sets), whereas deep interweaving IFs appear less clearly correlated.

\begin{table}[t]
 \centering
\subfloat[No interweaving]{
\begin{tabular}{rrr}
  \hline
 & 1 & 2 \\
  \hline
1 & 2901.87 &  \\
  2 & 2630.89 & 999.88 \\
  3 & 2931.02 & 233.08 \\
  4 & 2936.83 & 673.08 \\
  5 & 2909.98 & 876.57 \\
  6 & 2772.16 & 934.87 \\
  7 & 2303.80 & 958.02 \\
  8 & 1463.70 & 968.13 \\
  9 & 605.16 & 974.56 \\
  10 & 113.16 & 976.10 \\
   \hline
\end{tabular}
}
\qquad
\subfloat[Shallow interweaving]{
\begin{tabular}{rrr}
  \hline
 & 1 & 2 \\
  \hline
1 & 462.07 &  \\
  2 & 434.09 & 186.39 \\
  3 & 460.59 & 80.57 \\
  4 & 457.71 & 141.30 \\
  5 & 451.84 & 164.85 \\
  6 & 437.95 & 174.34 \\
  7 & 406.73 & 178.91 \\
  8 & 337.78 & 181.54 \\
  9 & 215.53 & 183.30 \\
  10 & 67.89 & 184.30 \\
   \hline
\end{tabular}
}
\qquad
\subfloat[Deep interweaving]{
\begin{tabular}{rrr}
  \hline
 & 1 & 2 \\
  \hline
1 & 8.56 &  \\
  2 & 10.81 & 8.69 \\
  3 & 8.48 & 10.92 \\
  4 & 8.55 & 9.00 \\
  5 & 8.79 & 8.46 \\
  6 & 9.33 & 8.25 \\
  7 & 10.38 & 8.19 \\
  8 & 12.36 & 8.17 \\
  9 & 16.07 & 8.14 \\
  10 & 22.07 & 8.18 \\
   \hline
\end{tabular}
}
\caption{Average IFs for factor loadings matrix $\facload$.}
\label{simstudy:meanIFs}
\end{table}

\begin{table}[t]
 \centering
 \begin{tabular}{lrrrr}
  \hline
  & $\fac_{1,1000}$ & $\fac_{2,1000}$ & $\hsv_{11,1000}$ & $\hsv_{12,1000}$ \\
  \hline
  No interweaving & 121.54 & 53.30 & 232.23 & 21.59 \\
  Shallow interweaving & 48.98 & 20.42 & 127.24 & 16.27 \\
  Deep interweaving &  3.79 & 3.76 & 5.44 & 5.85 \\
 \hline
\end{tabular}
\caption{Average IFs for the final factors $\fac_{jT}$ and their log-variances $\hsv_{m+j,T}$, $j \in \{1,2\}$.}
\label{simstudy:meanIFs2}
\end{table}

To provide a more complete picture, we list the inefficiency factors for all elements of $\facload$ for the various algorithms in Table~\ref{simstudy:meanIFs}, averaged over all $100$ runs. Note that shallow interweaving permits efficiency gains of around two- to eight-fold as opposed to the standard sampler, whereas deep interweaving delivers gains up to around $400$-fold.

It comes as no surprise that sampling (in)efficiencies of draws for the volatility parameters $\mupar_i$, $i \in \{1,\dots,m\}$ as well as $\phipar_i$ and $\sigmapar_i$, $i \in \{1,\dots,m+r\}$ are not affected substantially by this interweaving strategy, thus they are not reported here. It is however worth noting that the inefficiency of factor $\fmj{j}$ and factor log-variance draws $\hsvmi{m+j}$, $j \in \{1,\dots,r\}$ may be influenced by bad mixing of $\facload$. For illustration, IFs are reported for the final factors $\fac_{1T}$
and $\fac_{2T}$ and their log-variances $\hsv_{m+1,T}$ and $\hsv_{m+2,T}$
in Table~\ref{simstudy:meanIFs2}.

\begin{figure}[t]
 \centering
 \includegraphics[width=0.8\textwidth]{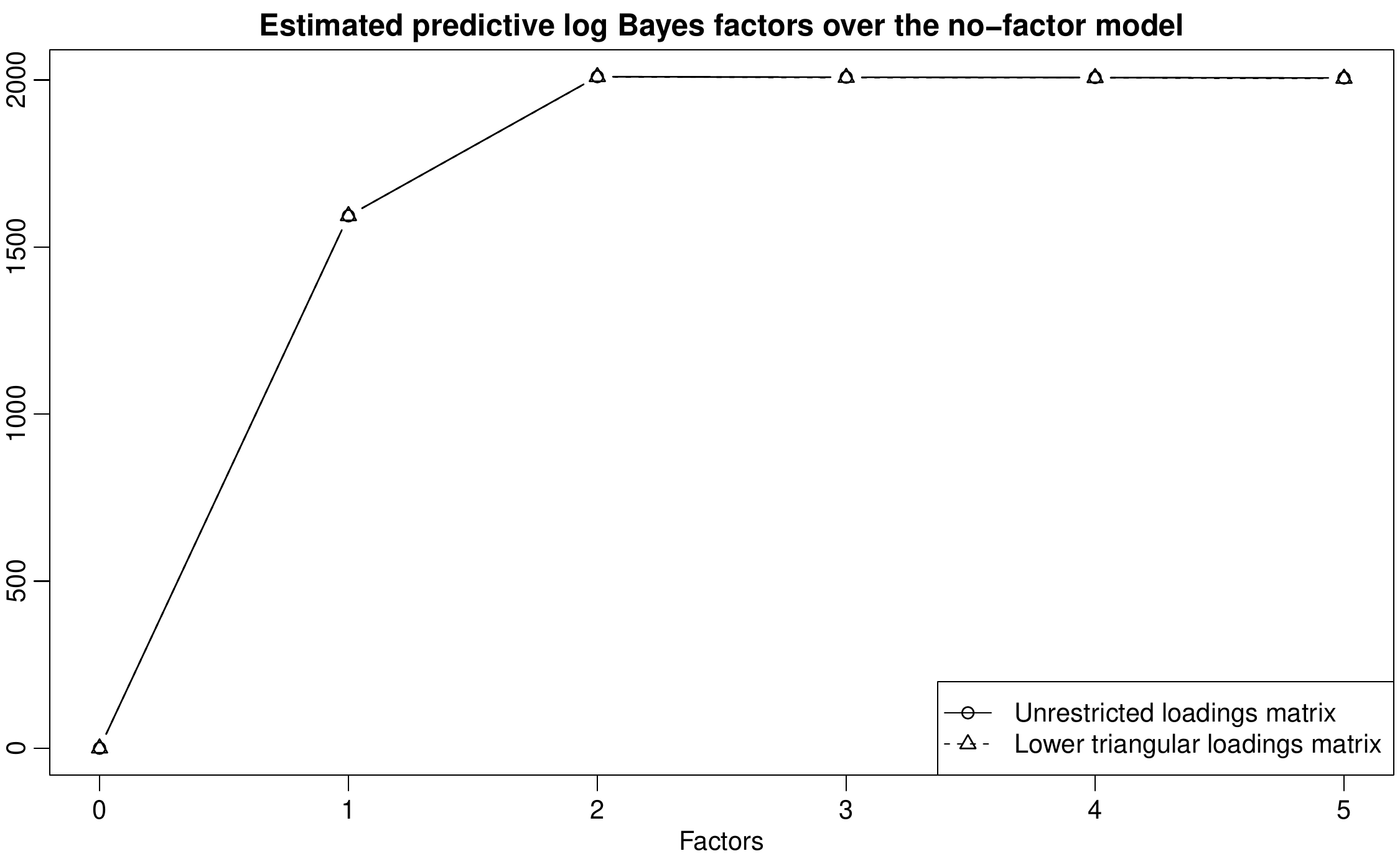}
 \caption{Log predictive Bayes factors in favor of the $r$-factor model over the no-factor model. The first $500$ returns in the data set are treated as prior information and log one-day-ahead predictive likelihoods are accumulated over the following $500$ days. Circles connected with a solid line indicate values obtained with completely unrestricted loadings matrices; triangles connected with dashed lines indicate values where the loadings matrix is restricted to be lower triangular.}
 \label{simdata:logpredlik}
\end{figure}

 To conclude the simulation exercise, we investigate predictive performance of under- and overfitting models through cumulative log predictive Bayes factors
 in Figure~\ref{simdata:logpredlik}; see \cite{kas:spa} for computational details. It stands out that the biggest predictive gain over a model that ignores
 contemporaneous correlations comes from introducing the first factor, i.e.~allowing for co-volatility through one common factor. Then, as expected, the
 second factor bumps the predictive score to its maximum. After that, it remains (almost) constant for three and more factors, irrespectively of whether the
  lower triangular restriction is enforced or not. This points out that underfitting models are severly worse in terms of prediction while overfitting models hardly
   suffer from the extra parameters introduced.

 Note that, similar to a scree plot in principal component analysis, Figure~\ref{simdata:logpredlik} can also be used as a graphical tool for finding
  the appropriate number of factors. For this exercise, we clearly find the true number of two factors.

\section{Application to Exchange Rate Data} \label{sec:app}

\begin{table}[t]
\centering
\begin{tabular}{rlrlrl}
  \hline
  AUD & Australia dollar  &
  CAD & Canada dollar &
  CHF & Switzerland franc \\
  CNY & China yuan renminbi &
  CZK & Czech R.\ koruna &
  DKK & Denmark krone \\
  GBP & UK pound &
  HKD & Hong Kong dollar &
  HRK & Croatia kuna \\
  HUF & Hungary forint &
  IDR & Indonesia rupiah &
  JPY & Japan yen \\
  KRW & South Korea won &
  MYR & Malaysia ringgit &
  NOK & Norway krone \\
  NZD & New Zealand dollar &
  PHP & Philippines peso &
  PLN & Poland zloty \\
  RON & Romania fourth leu &
  RUB & Russia ruble &
  SEK & Sweden krona \\
  SGD & Singapore dollar &
  THB & Thailand baht &
  TRY & Turkey lira \\
  USD & US dollar &
  ZAR &	South Africa rand \\
   \hline
\end{tabular}
\caption{Currency abbreviations.}
\label{abbrev}
\end{table}

In this section, we analyze exchange rates with respect to EUR. Data was obtained from the European Central Bank's Statistical Data Warehouse and ranges from April 1, 2005 to August 6, 2015. It contains $m=26$ (all which were available for this time frame) daily exchange rates on $2650$ days listed in Table~\ref{abbrev}. For further analysis, we thus use $T=2649$ demeaned log returns.
The data is displayed in Figure~\ref{exrates:dat} in Appendix~\ref{app:exchange}. Common ``stylized facts'' of financial time series are clearly visible; note e.g.\ the obvious volatility clustering during 2008 and 2009 and again throughout late 2014 and early 2015. To put the robustness of our sampler to the test, we use the data as-is, i.e.\ without excluding series containing extreme outliers such as the CHF spike on January 14, 2015 or the near collapse of RUB around December 16, 2014.

\subsection{Model Specification}


For selecting the number of factors in this application, it is important to keep in mind the primary purpose of the analysis.
Different sampling strategies are applied, depending on whether  identification of $\facload$ is of no concern (e.g.\ for covariance matrix prediction only), 
or whether identification  is instrumental for understanding  the unobserved, underlying factors.

\begin{figure}[t]
 \centering
 \includegraphics[width=0.8\textwidth]{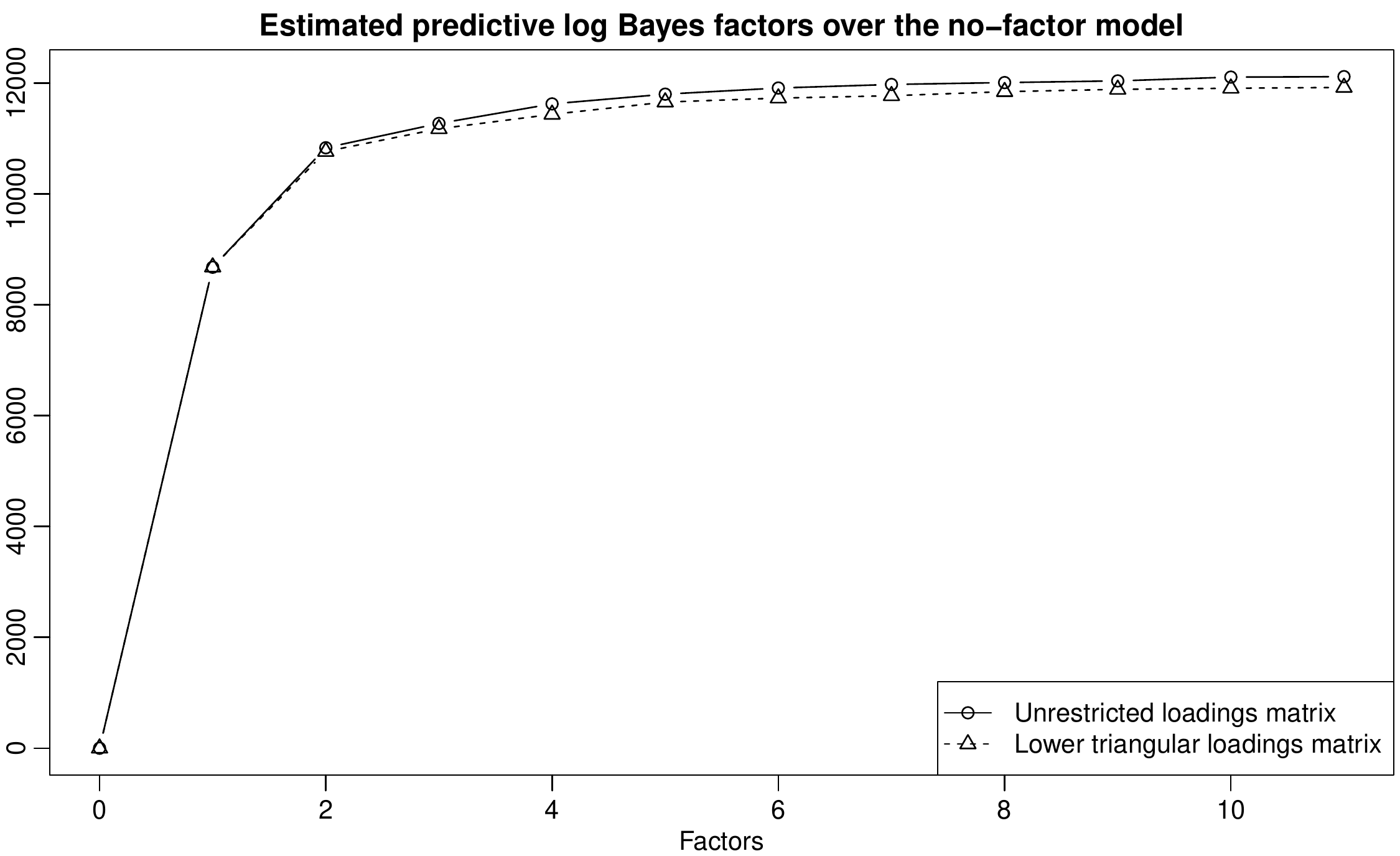}
 \caption{Log predictive Bayes factors in favor of the $r$-factor model over the no-factor model. The first $1000$ returns in the data set are treated as prior information and log one-day-ahead predictive likelihoods are accumulated over the following $1000$ days. Circles connected with a solid line indicate values obtained with completely unrestricted loadings matrices; triangles connected with dashed lines indicate values where the loadings matrix is restricted to be lower triangular.
 The component series are ordered alphabetically.}
 \label{exrates:logpredlik}
\end{figure}

For the first case, we  experimented with fitting unrestricted models to the exchange rates data. This implies that the method is completely invariant to series 
ordering and there are no model-implied ``leading factors'' as is usually the case.  
With respect to selecting the number of factors, we found that 
higher-order models without any restriction on the factor loadings
matrix yield higher marginal likelihoods and are thus recommended. Figure~\ref{exrates:logpredlik} illustrates this via log predictive Bayes factors.

If identification is warranted, an important step is the appropriate ordering of the variables, before 
 the usual  lower triangular structure is imposed on the factor loadings matrix 
  to guarantee mathematical identifiability, as outlined in Section~\ref{sec:ident}. 
 This, however, makes inference
on the factor loadings matrix dependent on the appropriate ordering of the variables. 
We exemplify this by the predictive Bayes factors for models where the component series are ordered alphabetically and the lower triangular structure 
is imposed on the first three series appearing in Table~\ref{abbrev}, namely AUD, CAD and CHF. As shown in Figure~\ref{exrates:logpredlik},
 for a given number of factors $r$,
 the  log predictive Bayes factors of the constrained models are consistently smaller than for the unrestricted models, 
  indicating that  a purely mathematical identifiability constraint  may  be in conflict with the data. 

Also for the constrained models, the log predictive Bayes factors are ever increasing for the exchange rate data.
However, it stands out
that the relative gain per additional factor is highest for few factors, flattening out quickly.
  Furthermore, draws of the factor loading matrix in these higher-order models are difficult to identify,  in particular, if 
   the lower triangular constraint is in conflict with the data    and  spurious
factors, i.e.\ factors which are significantly loaded on by only few series, are present
  \citep[see][for a detailed discussion of this issue in the static factor context]{fru-lop:par}. 
   Thus, to keep presentation feasible and to avoid spurious
factors, we restrict ourselves to a  model with $r=4$ factors for the following in-depth discussion.


One way to find constraints that are not in conflict with the data is to post-process the MCMC draws of an unrestricted sampler with $r=4$ 
  factors,  a method that has been  applied in  \citet{con-etal:bay} and \citet{ass-etal:bay}. Note that rather than reordering the variables 
   before imposing a lower triangular constraint,  we can choose three (out of the 26) currencies, and impose the  $r(r-1)/2=6$ zero restrictions on the
    corresponding factor loadings, see e.g.\ \citet{dun:not}.   While the choice of these currencies is not unique, inference 
    is robust to specific choices, as long as  the corresponding currencies    serve as  ``leaders'' for specific factors.
 This is exemplified by Figure~\ref{exrates:unrestricted} which displays the posterior median of the MCMC draws of the factor loadings  obtained from
 an unrestricted sampler with $r=4$.
 To solve column switching, the columns of $\facload$ are rearranged by the size of their maximum median loading.
 According to Figure~\ref{exrates:unrestricted} in Appendix~\ref{app:exchange}, USD is a definite candidate to lead factor one, PLN leads a second 
 factor, and AUD leads a third factor. Alternatively, identification could be based on any other currency strongly
 loading on factor 1 (such as HKD or CNY), in combination with HUF (instead of PLN) and NZD (instead of AUD).


Prior hyperparameters are the same as for the simulation study in Section~\ref{sec:sim}. A sensitivity analysis shows that none of the hyperparameter
choices turn out to be very influential in this particular application with the exception of the prior factor loadings variances $B_\Lambda$. These, however,
 are only important for the absolute scaling of the factors and do not notably influence the relative loadings sizes or
predictive Bayes factors.
We run each sampler for  $550\,000$ iterations, discard the first $50\,000$ draws as burn-in, leaving $500\,000$ which we use for posterior inference.
Even after this substantial amount of iterations, it is not clear that the sampler without interweaving has properly converged; we therefore omit its presentation.
 IFs from the interwoven samplers are presented in Table~\ref{ifexrates} in Appendix~\ref{app:exchange}.

Finally, we identify the signs of the loadings in the post-processing phase by investigating the MCMC draws. For each factor, the series whose posterior absolute loadings distribution is furthest away from zero is assigned a positive sign, the other loadings are aligned thereafter, see also Section~\ref{alg:sign}
in  Appendix~\ref{app:mcmc}.

\subsection{Posterior Factor Volatilities and their Loadings}

\begin{figure}[t!]
 \centering
 \includegraphics[width=\textwidth]{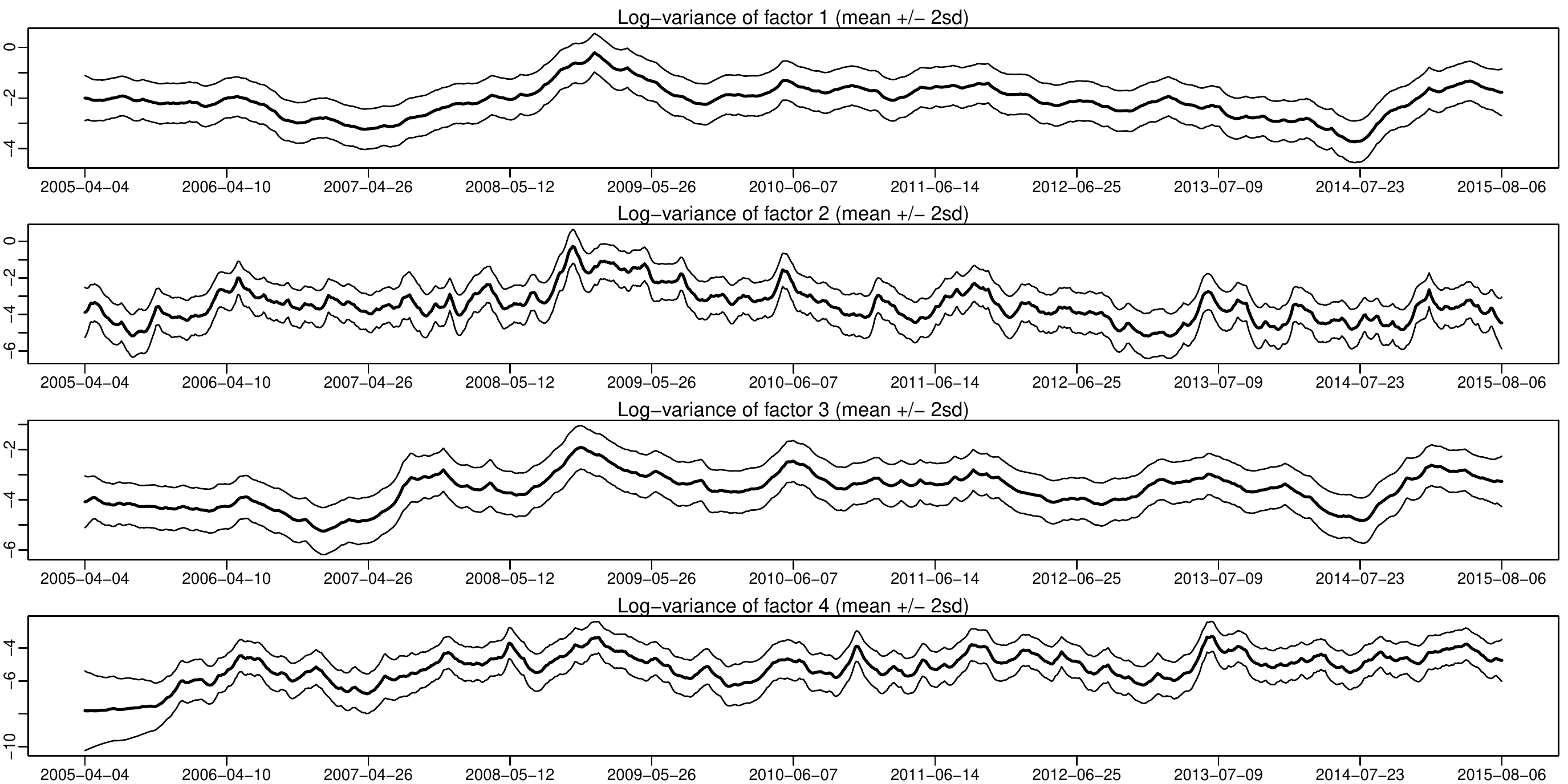}
 \caption{Marginal posteriors of the factor log-variances $\hsv_{m+j,t}, j = 1,\dots,4$ ($\text{mean} \pm 2 \times \text{sd}$).}
 \label{exrates:faclogvar}
\vspace*{\floatsep}
 \includegraphics[width=\textwidth]{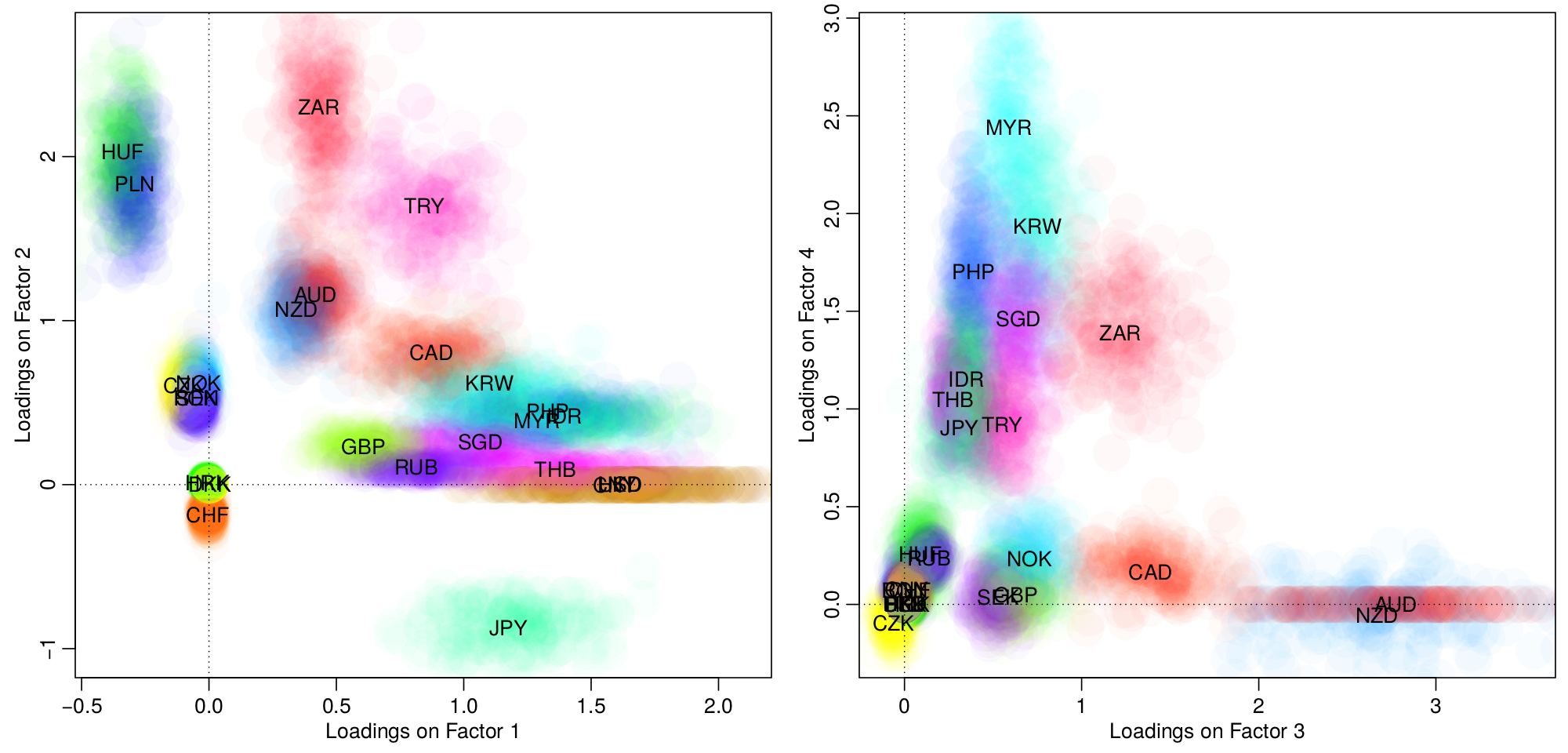}
\caption{Marginal posterior distribution of the factor loadings, visualized through arbitrarily colored scatterplots of MCMC draws.}
 \label{exrates:loadscatter}
\end{figure}

We begin by discussing the log-variances of the latent factors, visualized in Figure~\ref{exrates:faclogvar}, alongside the corresponding factor loadings
 whose marginal posterior distributions are depicted in Figure~\ref{exrates:loadscatter} and whose posterior means are listed in Table~\ref{exrates:loadmeans}.

\begin{table}[t]
\centering
{\small
\begin{tabular}{ccc}
\begin{tabular}{rrrrr}
  \hline
  & $\faccol{1}$ & $\faccol{2}$ & $\faccol{3}$ & $\faccol{4}$ \\
  \hline
%
  AUD & 0.418 & 1.156 & 2.772 & * \\
  CAD & 0.873 & 0.805 & 1.389 &  \\
  CHF &  & -0.184 &  &  \\
  CNY & 1.592 &  &  & 0.076 \\
  CZK & -0.099 & 0.605 &  &  \\
  DKK & 0.002 &  &  &  \\
  GBP & 0.605 & 0.230 & 0.627 &  \\
  HKD & 1.611 &  & 0.003 & 0.005 \\
  HRK &  &  &  &  \\
  HUF & -0.339 & 2.028 &  &  \\
  IDR & 1.395 & 0.419 & 0.347 & 1.153 \\
  JPY & 1.176 & -0.875 & 0.310 & 0.904 \\
  KRW & 1.100 & 0.617 & 0.750 & 1.935 \\
   \hline
\end{tabular}
&\hspace*{1cm}&
\begin{tabular}{rrrrr}
  \hline
  & $\faccol{1}$ & $\faccol{2}$ & $\faccol{3}$ & $\faccol{4}$ \\
  \hline
%
  MYR & 1.285 & 0.391 & 0.587 & 2.439 \\
  NOK &  & 0.619 & 0.704 &  \\
  NZD & 0.342 & 1.066 & 2.665 &  \\
  PHP & 1.330 & 0.449 & 0.389 & 1.702 \\
  PLN & -0.292 & 1.835 & * & * \\
  RON & -0.051 & 0.530 &  &  \\
  RUB & 0.813 & 0.104 & 0.138 & 0.237 \\
  SEK & -0.049 & 0.529 & 0.527 &  \\
  SGD & 1.065 & 0.260 & 0.642 & 1.463 \\
  THB & 1.358 & 0.092 & 0.273 & 1.049 \\
  TRY & 0.845 & 1.702 & 0.549 & 0.920 \\
  USD & 1.614 & * & * & * \\
  ZAR & 0.431 & 2.303 & 1.219 & 1.390 \\
   \hline
\end{tabular}
\end{tabular}}
\caption{Posterior means of $p(\facload|\ym)$, in alphabetical order. Blank entries signify that the respective marginal distribution is not bound away from zero with at least 99\% posterior probability.
Starred entries are those which have been set to zero a priori.}
\label{exrates:loadmeans}
\end{table}

 The first factor can clearly be interpreted as the USD-driven one, as the pegged triplet USD, CNY and HKD loads very highly on this factor, alongside many other currencies. Its volatility is generally very smooth, rising in the aftermath of the 2008 financial crisis and going down again after 2009; a second increase can be seen in the second half of 2014, possibly in connection with the Greek government-debt crisis.
 Factor~2's log-variance appears slightly less persistent and more volatile, it is driven by ZAR, the only African currency in the sample, alongside Eastern Europe's / Southwestern Asia's HUF, PLN and TRY. Interestingly, JPY loads negatively on this factor.
 The third factor shows a similar overall pattern as the first. The highest loading series for this factor are AUD and NZD, emphasizing the Trans-Tasman relations. Other commodity currencies such as ZAR and CAD also load highly on this factor.
 Factor~4 is clearly driven by the currencies of the Tiger Cub economies such as MYR, KRW, PHP and SGD.

\subsection{Posterior Volatilities and Correlations}


In order not to overload the graphical displays used to visualize the results of the analysis, we display the results for a two-year period only for the rest of this section. More specifically, we look at the years 2008 and 2009, covering the most volatile span during the financial crisis. Irrespectively of that, the full data set has been used for estimation and other time spans could be displayed analogously. 

\begin{figure}[p]
 \centering
 \includegraphics[width=\textwidth]{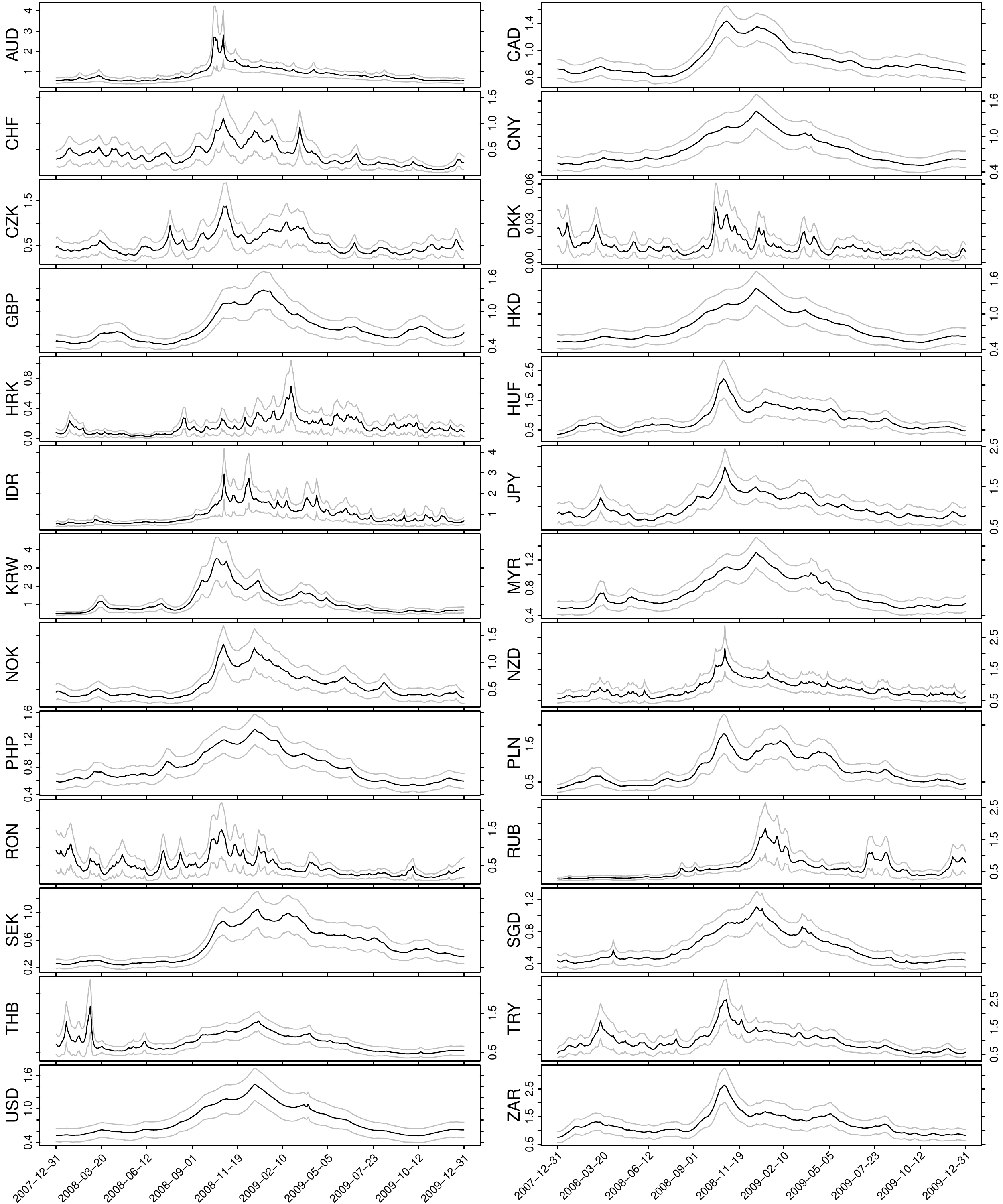}
 \caption{Posterior volatilities of exchange rate log returns with respect to EUR, from the last day of 2007 until the last day of 2009 ($\text{mean} \pm 2 \times \text{sd}$).}
 \label{exrates:vols}
\end{figure}

We start out by visualizing the marginal posterior means of univariate volatilities for all $26$ currencies in Figure~\ref{exrates:vols} from the last day of 2007 until the last day of 2009. Series such as DKK or HRK are (very) closely pegged to EUR and unsurprisingly show very low volatility throughout the crisis. Other European currencies (CHF\footnote{It is interesting to note that the Swiss franc stays comparably stable from a EUR perspective throughout 2007-2009. Very differently during summer 2011, where CHF shows atypical and very high volatility until the Swiss Central Bank sets the minimum exchange rate at CHF 1.20 per EUR 1 on September 6.}, RON, SEK, CZK) follow suit. Tiger Cub economies such as PHP, HKD, THB, and MYR align very closely with USD and CNY. The most volatile currencies during this period are KRW, ZAR, IDR, JPY, and also TRY, followed by NZD, AUD, and CAD. Overall, it stands out that even though some series-specific ups and downs can be spotted, a common trend is clearly visible.

\begin{figure}[t!]
 \centering
  \includegraphics[width=\textwidth]{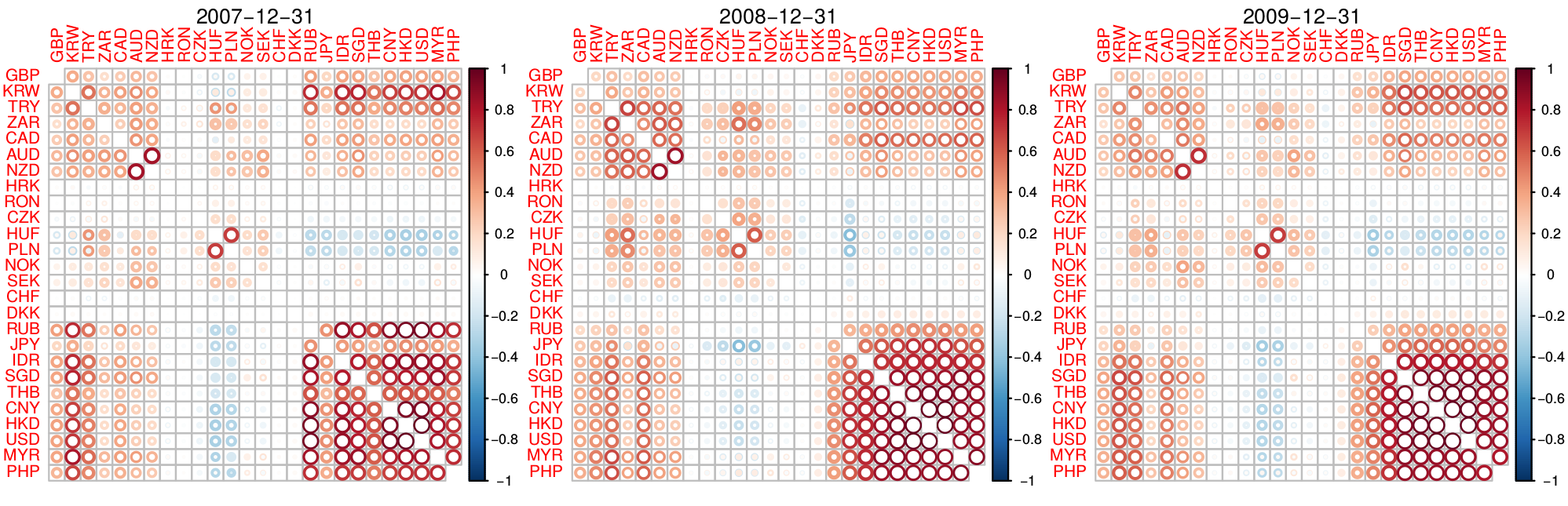}
  \caption{Posterior correlation matrices on the last trading days of 2006, 2007, and 2008. For each element, the size of the outer/inner circle is determined by the posterior mean plus/minus two posterior standard deviations, thereby indicating posterior uncertainty. Color and opacity are determined by the posterior mean. The remaining days are visualized in a video to be found online at \protect\url{https://vimeo.com/212887492}.}
 \label{exrates:cormat}
\end{figure}

Next, implied correlation matrices are displayed in Figure~\ref{exrates:cormat}, exemplified for the last day of 2007, 2008, and 2009. 
Additionally to displaying the mean posterior pairwise correlations (at the given dates) via color and shading, these plots visualize posterior uncertainty; the outer and inner circles' sizes correspond to $\text{posterior mean} \pm 2\text{ standard deviations}$, respectively. The images were generated using the \proglang{R} package \pkg{corrplot} \citep{r:corrplot}; its option \code{hclust} (hierarchical clustering) was used for ordering the series to emphasize the blocks of currencies.

To further illustrate variability over time, we determine the posterior means of the pairwise time-varying correlations of USD against the other currencies which are  plotted in Figure~\ref{exrates:USDcor} in Appendix~\ref{app:exchange}. As was to be expected, correlations of CNY and HKD with USD are almost always very close to one; IDR, THB, SGD, MYR, and PHP show rather high correlation throughout. The correlation between USD and RUB on the other hand falls from around $0.9$ in early 2008 to around $0.4$ in late 2009, whereas THB moves in the opposite direction; its correlation with USD is around $0.5$ at the beginning of the time window and increases quickly to around $0.9$. Eastern European non-euro currencies, in particular PLN and HUF, appear to be slightly negatively correlation with USD throughout the entire period.

\section{Conclusion  and Outlook} \label{sec:con}

Estimating time-varying (dynamic) covariance and correlation matrices of financial and economic time series constitutes a current and active area of research. One of the main challenges thereby is the curse of dimensionality, i.e.\ the fact that the number of elements of these matrices grows quadratically with the number of observed series. We address this issue by imposing a low-dimensional latent factor structure where the factors are allowed to exhibit stochastic volatility and thereby govern co-movement of volatility over time. To conduct reliable statistical inference, we propose novel Bayesian MCMC algorithms which exploit the model-inherent identifiability constraints. By interweaving different (but mathematically equivalent) parameterizations, the proposed strategies substantially improve mixing of draws obtained from the posterior distribution, in particular for the factor loadings matrix. The method proposed is fully automatic in the sense that the end-user is not required to manually adjust any tuning parameters. 

In an extensive case study discussing exchange rates with respect to EUR we show that the algorithm plays well with real-world data that exhibits a fair degree of outliers (e.g.\ CHF, RUB) which are captured through the idiosyncratic stochastic volatility components. The model structure allows for a covariance decomposition in four interpretable factors (USD/CNY driven, Eastern Europe, commodity currencies, Tiger Cub economies). These, alongside the idiosyncratic volatilities, drive the dynamics of the joint correlation structure. The pairwise correlations with USD range from ``almost perfect'' (CNY, HKD) over ``hardly existent'' (CHF, HRK) to ``slightly negative'' (PLN, HUF) with a varying and time-dependent degree of variability.

Concerning extensions of the model, we point out that due to the modular nature of MCMC, all the ideas of this paper can be straightforwardly generalized
 to models that independently model the mean, be it through a simple nonzero mean vector, a local level model, external regressors, or via (vector) autoregressive
 processes. For models where the level of the returns explicitly depends on the (co-)volatilities \citep[\emph{volatility-in-mean}-type-effects, see e.g.][]{cha:sto}
 or the returns are assumed to be correlated with the (co-)volatilities \citep[\emph{leverage}-type-effects, see e.g.][]{ish-omo:por}, more involved
 estimation methods are required; this unfortunately places their discussion outside the scope of this paper. Nevertheless, due to the growing number of
 successful applications of interweaving methods in different contexts, there is good reason to hope for similar effects when they are used for these type
  of factor SV model extensions.

\bigskip
\begin{center}
{\large\bf SUPPLEMENTARY MATERIAL}
\end{center}
\begin{description}
\item[\textsc{Web Appendix}] containing (\ref{app:mcmc}) details of the various sampling steps of Algorithm~\ref{facsvalg} and details concerning the \proglang{R} package \pkg{factorstochvol}, (\ref{app:sim}) the data generating parameter values for the simulation study in Section~\ref{sec:sim}, and  (\ref{app:exchange}) further results for the  exchange rate data discussed in Section~\ref{sec:app}. (.pdf file)
\item[\textsc{Video}] displaying the time-varying conditional correlation matrix distribution for the full data set (cf.\ Figure~\ref{exrates:cormat}). Available at \url{https://vimeo.com/212887492}.
(.avi file)

\item[\textsc{\proglang{R}-package}] \pkg{factorstochvol}, version 0.8.3, containing code to run the samplers described in the paper. Available at \url{https://cran.r-project.org/package=factorstochvol}.
(GNU zipped tar file)


\item[\textsc{Replication code}] for the results in the paper, including the exchange rate data. (GNU zipped tar file)

\end{description}


\newpage \appendix
\setcounter{equation}{0}

\renewcommand{\thetable}{A.\roman{table}}
\renewcommand{\thefigure}{A.\arabic{figure}}
\renewcommand{\thesection}{A.\arabic{section}}
\renewcommand{\theequation}{A.\arabic{equation}}
\numberwithin{table}{section}
\numberwithin{figure}{section}
\numberwithin{equation}{section}


\begin{appendix}

\thispagestyle{empty}

\begin{center}
 {\LARGE \bf Web Appendix to ``Efficient Bayesian Inference\\[-.4em]for Multivariate Factor Stochastic Volatility\\[-.4em]Models''\\[5pt]
\vspace{1cm} }


\textsc{Gregor Kastner} \\ {\it  Department of Finance, Accounting and Statistics\\WU Vienna University of Economics and Business, Austria}\\
 \url{gregor.kastner@wu.ac.at}
\\[10pt]
\textsc{Sylvia Fr\"uhwirth-Schnatter}\\ {\it  Department of Finance, Accounting and  Statistics\\WU Vienna University of Economics and Business, Austria}\\ \url{sfruehwi@wu.ac.at} \\[10pt]

\textsc{Hedibert Freitas Lopes}\\ {\it  Insper, Brazil}\\
  \url{hedibertfl@insper.edu.br}\\[10pt]

\vspace{0.5cm}

\today \\[5pt]

\end{center}

\setcounter{page}{1}

\hspace*{1cm}

\begin{abstract}
 \noindent This document contains supplementary material for the paper ``Efficient Bayesian Inference for Multivariate Factor Stochastic Volatility Models''. It encloses (\ref{app:mcmc}) details of the various sampling steps of Algorithm~\ref{facsvalg} and details concerning the \proglang{R} package \pkg{factorstochvol}, (\ref{app:sim}) the data generating parameter values for the simulation study in Section~\ref{sec:sim}, and (\ref{app:exchange}) further results for the  exchange rate data discussed in Section~\ref{sec:app}.
\end{abstract}

\clearpage

\section{Details on MCMC Sampling}  \label{app:mcmc}

\subsection{Details on Updating the Volatilities} \label{alg:univ}
The latent equations  
\begin{eqnarray}
\hsv_{it}&=& (1-\phipar_i)\mupar_i +  \phipar_i \hsv_{i,t-1} + \sigmapar_i \etat_{it}, \qquad i=1, \ldots,m,
\label{fac3app}  \\
\hsv_{m+j,t}&=& \phipar_{m+j} \hsv_{m+j,t-1} + \sigmapar_{m+j} \etat_{m+j,t}, \qquad j=1, \ldots, r, \label{fac4app}
\end{eqnarray}
 are combined with the (augmented) observation equation (\ref{fac4}),
yielding
\begin{eqnarray}
 \log (y_{it} -  \facrow{i} \facm_t)^2  &=&  \hsv_{it} + \log \error_{it}^2, \qquad i=1,\ldots,\dimy, \label{facob1a} \\
\log \fac_{jt}^2  &=&  \hsv_{m+j,t} + \log  \errorstate_{jt}^2, \qquad j=1,\ldots,\nfactrue. \label{facob2a}
 \end{eqnarray}
Due to the modular nature of MCMC methods, updating the latent log-variances $\hsvm$ and the corresponding parameters appearing in (\ref{fac3app}) and
(\ref{fac4app}) amounts to $m+r$ independent updates with augmented data appearing on the left hand side of (\ref{facob1a}) and (\ref{facob2a}).

Each of these $m+r$ models is a univariate stochastic volatility (SV) model as introduced by \citet{tay:fin}.
References about its efficient Bayesian estimation include \citet{jac-etal:bayJBES, she:par, she-pit:lik, kim-etal:sto, omo-etal:sto, str-etal:par, mcc-etal:sim, kas-fru:anc, she-nea:eff}.
Consequently, the substantial amount of research on this matter which has emerged in the last two decades can directly be applied. In particular,
we follow \citet{kas-fru:anc} and simply use the implementation in the \proglang{R} \citep{r:r} package \pkg{stochvol} \citep{r:sto} as a ``plug-in'' for the factor SV sampler discussed in this paper (set \code{dontupdatemu} to \code{TRUE} for sampling factor volatilities). The SV update in \pkg{stochvol} can be accessed from \proglang{R} through the functions \code{svsample} and \code{svsample2}, whereas the latter is a stripped-down version of the former that omits input checking and post-processing. Moreover, in order to maximize execution speed when large models are to be fitted, \proglang{C/C++} level access to the core \code{update} function is provided. Note that the number of function calls may be in the range of billions and more, thus even tiny costs for code interpretation can quickly accumulate. The \proglang{R} level interface was employed for prototyping and proof-of-concept implementations \citep[cf.][]{kas-etal:ana}. For the results reported in this paper, direct access to \code{update} was used. A detailed description of this procedure is given in \citet{kas:dea} and the package manual.



\subsection{Details on Sampling the Loadings} \label{alg:facload}

Because $y_{it}\sim \Normal{\facrow{i} \facm_t, \e^{h_{it}}}$, sampling the loadings
conditionally on $\facm$
constitutes a Bayesian regression problem with heteroscedastic errors.
Letting $\tilde r_i$ denote the number of unrestricted elements in row $i$, rewriting yields
\[
\tilde\ym_i \sim \Normult{T}{\bm{X}_i \facrow{i}',\unit{T}},
\]
where $\tilde \ym_{i}=(y_{i1}\e^{-h_{i1}/2},\dots,y_{iT}\e^{-h_{iT}/2})'$ denotes the $i$th normalized observation vector and
\[
\bm{X}_i=
\begin{bmatrix}
\fac_{11}\e^{-h_{i1}/2}& \cdots & \fac_{\tilde r 1}\e^{-h_{i1}/2} \\
\vdots & & \vdots \\
\fac_{1T}\e^{-h_{iT}/2}& \cdots & \fac_{\tilde r T}\e^{-h_{iT}/2}
\end{bmatrix}
\]
is the $T\times \tilde r_i$ design matrix. Thus, independently for each $i$, sampling from $\facrow{i}|\facm, \ymi{i},\hsvmi{i}$ is achieved by
performing a Gibbs-update from
\[
\facrow{i}'|\facm, \ymi{i},\hsvmi{i}, \sim \Normult{\tilde r}{\bm{b}_{iT}, \bm{B}_{iT}},
\]
with $\bm{B}_{iT}=(\bm{X}_i'\bm{X}_i + B_\load^{-1}  \identy{\tilde r_i})^{-1}$  
and $\bm{b}_{iT}=\bm{B}_{iT}\bm{X}_i'\tilde \ym_i$.

\subsection{Details on Sampling the Factors}  \label{alg:samfac}

Sampling the factors $\facm_{t}$ for each $t=1,\ldots, T$ conditionally on the factor loadings $\facload$ and the volatilities $\hsvm_{t}$, i.e.\ from $\facm_{t}|\facload,\ym_{t},\hsvm_{t}$,
is again a standard Bayesian regression problem. We have
\[
\tilde\ym_{t} \sim \Normult{m}{\bm{X}_t\facm_{t},\unit{m}},
\]
where $\tilde \ym_{t}=(y_{1t}\e^{-h_{1t}/2},\dots,y_{mt}\e^{-h_{mt}/2})'$ denotes the normalized observation vector at time $t$
and
\[
\bm{X}_t=\begin{bmatrix}
\load_{11}\e^{-h_{1t}/2} & \cdots & \load_{1r}\e^{-h_{1t}/2}\\
\vdots&&\vdots\\
\load_{m1}\e^{-h_{mt}/2} & \cdots & \load_{mr}\e^{-h_{mt}/2}\\
\end{bmatrix},
\]
is the $m\times r$ design matrix. Independently for each $t$, the posterior is consequently given by
\[
\facm_{t}|\facload,\ym_{t},\hsvm_{t} \sim \Normult{r}{\bm{b}_{mt}, \bm{B}_{mt}},
\]
where $\bm{B}_{mt}=(\bm{X}_t'\bm{X}_t + \Vm_t(\hsvm^V_t)^{-1})^{-1}$ and $\bm{b}_{mt}=\bm{B}_{mt}(\bm{X}_t'\tilde \ym_t)$.

\subsection{Sign Identification}  \label{alg:sign}

For sensible interpretation of the factor loadings their signs have to be identified. If done a posteriori, this requires the selection of one series per factor whose loadings distribution is sufficiently bound away from zero. For the paper at hand, we simply investigate the absolute values of the posterior draws from the loadings distribution. For each factor, the series whose smallest absolute MCMC draw is largest gets assigned a positive sign which in turn implies the signs for all other loadings on this factor. In other words, letting $K$ denote the number of MCMC draws after burn-in, in order to identify factor $j = 1,\dots,r$, we assign a positive sign to the posterior distribution of $|\load_{ij}|$, where $i$ is chosen to be
\[
 \argmax_{i \in \{1,\ldots,m\}} \left(\min_{k \in \{1,\ldots,K\}} \left|\load_{ij}^{(k)}\right|\right).
\]
Subsequently, we align the other loadings on factor $j$ accordingly.
To achieve this behavior in \pkg{factorstochvol}, use \code{signident} with \code{method = "maximin"}. In order to simply use the leading factors to identify the loadings, use \code{signident} with \code{method = "diagonal"}.
For the exchange rate data, the sign of factor 1 is identified via USD, factor 2 via ZAR, factor 3 via AUD, and factor 4 via MYR.

\subsection{A Note on Implementation} \label{sec:impl}

High-dimensional models, in particular models with many latent variables, pose a non-negligible computational challenge to those aiming for efficient MCMC implementations.
In principle, each individual MCMC step can be straightforwardly computed in parallel. In practice, however, doing so is only useful in shared memory environments (e.g.\ through multithreading/multiprocessing) as the increased communication overhead in distributed memory environments easily outweighs the speed gains.
Apart from within-steps-parallelization, MCMC is of intrinsically iterative nature in the sense that posterior draws are generated conditionally on older draws. Thus, they cannot be parallelized straightforwardly and call for compiled and optimized programming languages to avoid the cost of code interpretation at every iteration. Moreover, memory access needs to be optimized, as large amounts of latent variable draws must be stored either temporary (if required only for the next conditional draws) or more permanently (if required for direct posterior inference). In this paper, we tackle the computational burden by using high-performance \proglang{C} and \proglang{C++} code, interfaced to \proglang{R} via \pkg{RcppArmadillo} \citep{edd-san:rcp}. Additionally to providing an interface between \proglang{R} and \proglang{C++}, \pkg{RcppArmadillo} also accommodates fast linear algebra routines by means of the \pkg{Armadillo} library \citep{san:arm}. For ease-of-use, all code is bundled in the \proglang{R} package \pkg{factorstochvol} \citep{kas:fac} which is available on the Comprehensive \proglang{R} Archive Network (CRAN).

\section{Simulation Study} \label{app:sim}

The data generating parameter values for the simulation study in Section~4 are listed in Table~\ref{simstudy:truevals}.

\begin{table}[t!]
\centering
\subfloat[Factor loadings]{
\begin{tabular}{rrr}
  \hline
$\facloadtrue$ & 1 & 2 \\
  \hline
1 & 1.00 & \\
  2 & 0.90 & 1.00 \\
  3 & 0.80 & 0.10 \\
  4 & 0.70 & 0.20 \\
  5 & 0.60 & 0.30 \\
  6 & 0.50 & 0.40 \\
  7 & 0.40 & 0.50 \\
  8 & 0.30 & 0.60 \\
  9 & 0.20 & 0.70 \\
  10 & 0.10 & 0.80 \\
   \hline
\end{tabular}
}
\qquad
\subfloat[Idiosyncratic volatility parameters]{
\begin{tabular}{rrrr}
  \hline
 & $\mupartrue$ & $\phipartrue$ & $\sigmapartrue$ \\
  \hline
1 & -2.00 & 0.80 & 0.60 \\
  2 & -1.90 & 0.82 & 0.55 \\
  3 & -1.80 & 0.84 & 0.50 \\
  4 & -1.70 & 0.86 & 0.45 \\
  5 & -1.60 & 0.88 & 0.40 \\
  6 & -1.50 & 0.90 & 0.35 \\
  7 & -1.40 & 0.92 & 0.30 \\
  8 & -1.30 & 0.94 & 0.25 \\
  9 & -1.20 & 0.96 & 0.20 \\
  10 & -1.10 & 0.98 & 0.15 \\
   \hline
\end{tabular}
}
\qquad
\subfloat[Factor volatility parameters]{
\begin{tabular}{rrr}
  \hline
 & $\phipartrue$ & $\sigmapartrue$ \\
  \hline
  1  & 0.99 & 0.10 \\
  2  & 0.95 & 0.30 \\
   \hline
\end{tabular}
}
\caption{Data generating values.}
\label{simstudy:truevals}
\end{table}

\section{Application to Exchange Rate Data} \label{app:exchange}

\begin{figure}[t!]
 \centering
 \includegraphics[height=.93\textheight]{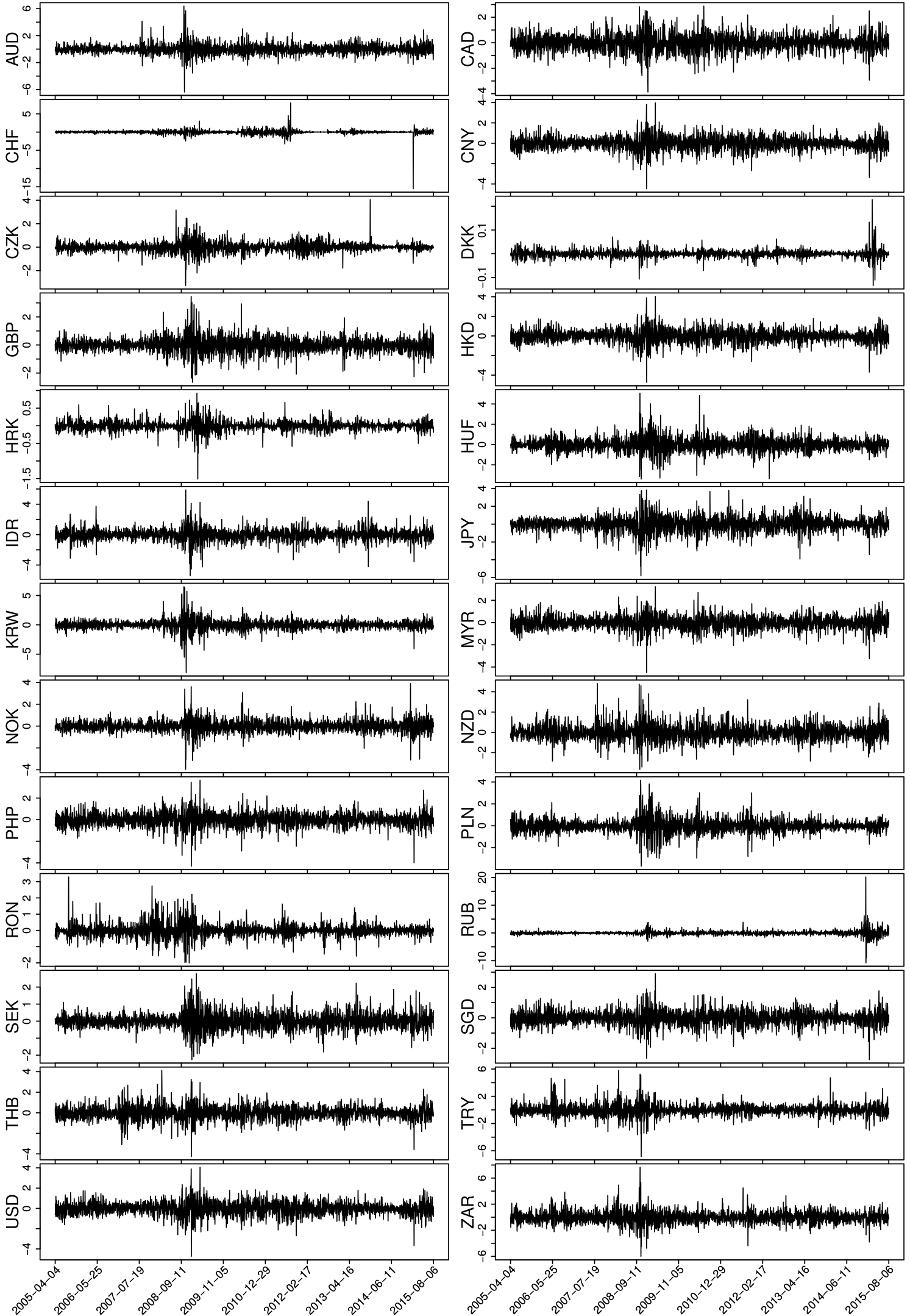}
 \caption{Demeaned log returns of EUR exchange rates.}
 \label{exrates:dat}
\end{figure}

\begin{figure}[h]
 \begin{center}
 \includegraphics[width=\textwidth]{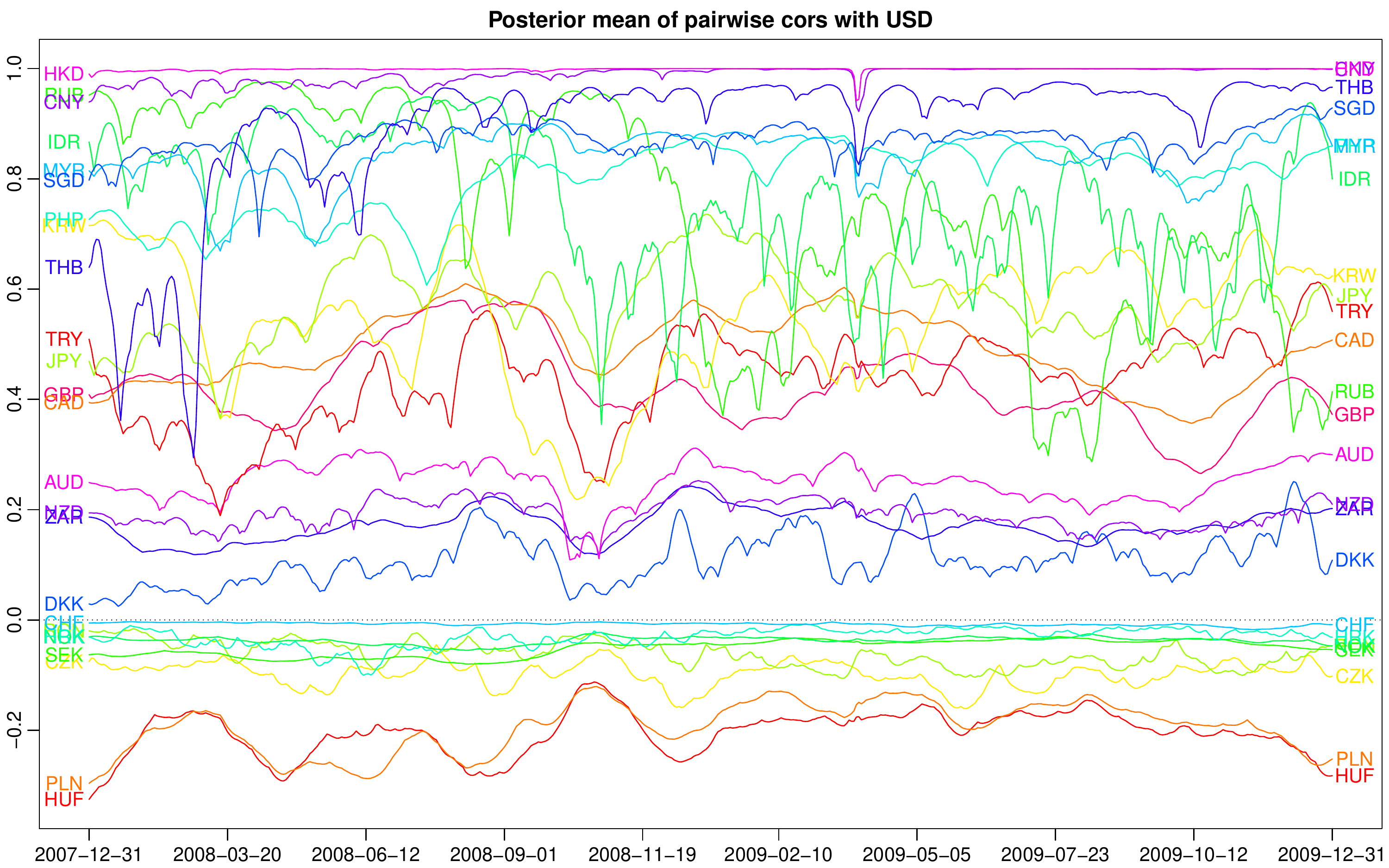}
 \caption{Posterior means of pairwise correlations with USD.}
 \label{exrates:USDcor}
 \end{center}
\end{figure}

\begin{figure}[p]
 \begin{center}
 \includegraphics[page=1,width=.92\textwidth]{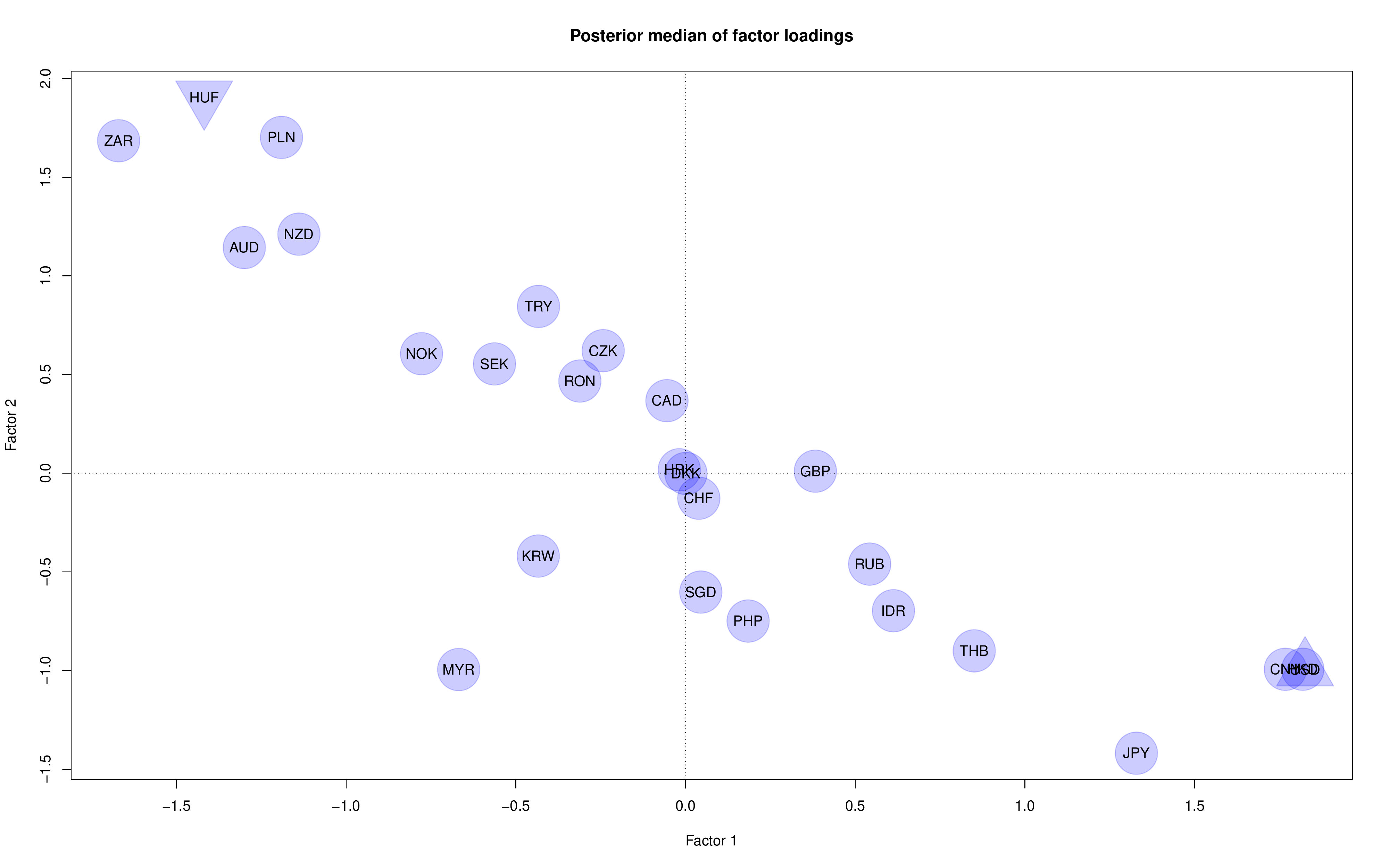}
 \includegraphics[page=2,width=.92\textwidth]{loadings3_unrestricted.pdf}
 \caption{Posterior medians of the factor loadings in the unrestricted model. The factors have been ordered according to their average posterior median loadings. An up-pointing triangle means that this series was used for sign-identification in the horizontal direction, a down-pointing triangle means it was used for sign-identification in the vertical direction.}
 \label{exrates:unrestricted}
 \end{center}
\end{figure}

This section provides additional results for exchange rate data analyzed in Section~5.
Figure~\ref{exrates:dat} depicts the raw data.
The pairwise correlations of USD against the other currencies as implied by the model discussed in detail in the main paper are plotted in Figure~\ref{exrates:USDcor}.
Inefficiency factors for posterior draws from the factor loadings using the interwoven samplers, i.e.\ deep and shallow interweaving, are presented in Table~\ref{ifexrates}.
Finally, Figure~\ref{exrates:unrestricted} visualizes the posterior medians of the factor loadings in a $4$-factor model without any restrictions on the factor loadings matrix. The factor ordering has been determined through a post-processing procedure; factors are sorted according to their average median posterior loadings.

\begin{table}[htp]
\centering
\subfloat[Shallow interweaving]{
\begin{tabular}{rrrrr}
  & $\faccol{1}$ & $\faccol{2}$ & $\faccol{3}$ & $\faccol{4}$ \\
  \hline
%
AUD & 414 & 588 & 901 & * \\ 
  CAD & 704 & 608 & 842 & 25 \\ 
  CHF & 8 & 203 & 38 & 25 \\ 
  CNY & 784 & 12 & 21 & 128 \\ 
  CZK & 207 & 770 & 28 & 19 \\ 
  DKK & 99 & 7 & 23 & 6 \\ 
  GBP & 719 & 176 & 717 & 9 \\ 
  HKD & 786 & 40 & 74 & 72 \\ 
  HRK & 11 & 10 & 13 & 5 \\ 
  HUF & 399 & 861 & 58 & 30 \\ 
  IDR & 770 & 525 & 386 & 609 \\ 
  JPY & 749 & 661 & 129 & 228 \\ 
  KRW & 752 & 555 & 669 & 812 \\ 
  MYR & 766 & 420 & 582 & 911 \\ 
  NOK & 28 & 618 & 665 & 28 \\ 
  NZD & 315 & 526 & 881 & 56 \\ 
  PHP & 784 & 557 & 417 & 813 \\ 
  PLN & 388 & 903 & * & * \\ 
  RON & 70 & 710 & 26 & 14 \\ 
  RUB & 780 & 224 & 189 & 146 \\ 
  SEK & 39 & 630 & 600 & 11 \\ 
  SGD & 785 & 356 & 803 & 898 \\ 
  THB & 780 & 82 & 483 & 753 \\ 
  TRY & 694 & 824 & 342 & 230 \\ 
  USD & 786 & * & * & * \\ 
  ZAR & 358 & 855 & 608 & 340 \\ 
   \hline
\end{tabular}
}
\qquad
\subfloat[Deep interweaving]{
\begin{tabular}{rrrrr}
  & $\faccol{1}$ & $\faccol{2}$ & $\faccol{3}$ & $\faccol{4}$ \\
  \hline
%
AUD & 32 & 49 & 29 & * \\
  CAD & 26 & 40 & 27 & 17 \\
  CHF & 6 & 45 & 28 & 14 \\
  CNY & 27 & 9 & 11 & 27 \\
  CZK & 18 & 47 & 10 & 9 \\
  DKK & 10 & 7 & 10 & 5 \\
  GBP & 25 & 24 & 24 & 9 \\
  HKD & 27 & 36 & 35 & 34 \\
  HRK & 5 & 6 & 6 & 5 \\
  HUF & 27 & 47 & 50 & 16 \\
  IDR & 26 & 39 & 23 & 30 \\
  JPY & 26 & 39 & 40 & 24 \\
  KRW & 26 & 38 & 26 & 31 \\
  MYR & 26 & 43 & 28 & 34 \\
  NOK & 16 & 38 & 24 & 13 \\
  NZD & 34 & 46 & 28 & 51 \\
  PHP & 26 & 39 & 23 & 32 \\
  PLN & 26 & 49 & * & * \\
  RON & 16 & 42 & 26 & 10 \\
  RUB & 26 & 28 & 18 & 22 \\
  SEK & 15 & 39 & 24 & 12 \\
  SGD & 27 & 42 & 29 & 33 \\
  THB & 26 & 25 & 25 & 32 \\
  TRY & 26 & 42 & 29 & 23 \\
  USD & 27 & * & * & * \\
  ZAR & 27 & 43 & 26 & 24 \\
  \hline
\end{tabular}
}
\caption{Estimated inefficiency factors for posterior draws of the factor loadings.}
\label{ifexrates}
\end{table}

\eject
\cleardoublepage

\end{appendix}

\bibliography{ref}




\end{document}